\documentclass[
reprint,
superscriptaddress,
amsmath,amssymb,
aps,
prl,
]{revtex4-1}

\usepackage[latin1]{inputenc}
\usepackage{amsmath}
\usepackage{amsfonts}
\usepackage{hyperref}
\usepackage{amssymb}
\usepackage{graphicx}
\usepackage{hyperref}
\usepackage{colortbl}
\usepackage{subfigure} 
\usepackage{gensymb}
\usepackage{svg}    

\usepackage[T1]{fontenc}
\usepackage{mathptmx}
\usepackage{bbold}
\usepackage{scalerel}
\usepackage{xcolor}
\usepackage{mathtools}
\usepackage{epstopdf}
\usepackage{changes}

\newcommand{\hyref}[1]{\hyperref[#1]{\ref{#1}}}

\newcommand{\orange}[1]

\newcommand{\ktea}{k$_{\rm B}$T/e\AA}
\newcommand{\xy}{X$^{2+}$Y$^{2-}$~}
\newcommand{\angstrom}{\text{\AA}}

\begin{document}

\title{Ion Selectivity in Uncharged Tapered Nanoslits through Heterogeneous Water Polarization  }
\author{Tim~E. Veenstra}
\email{t.e.veenstra@uu.nl}
    \affiliation{Soft Condensed Matter \& Biophysics, Debye Institute for Nanomaterials Science, Utrecht University, Princetonplein 1, 3584 CC Utrecht, The Netherlands}
    \affiliation{Institute for Theoretical Physics, Utrecht University,  Princetonplein 5, 3584 CC Utrecht, The Netherlands}
 \author{Gerardo Campos-Villalobos}
	\affiliation{Soft Condensed Matter \& Biophysics, Debye Institute for Nanomaterials Science, Utrecht University, Princetonplein 1, 3584 CC Utrecht, The Netherlands}
    \affiliation{CNR-ISC and Department of Physics, Sapienza University of Rome, p.le A. Moro 2, 00185 Roma, Italy}
 \author{Giuliana Giunta}
	\affiliation{Soft Condensed Matter \& Biophysics, Debye Institute for Nanomaterials Science, Utrecht University, Princetonplein 1, 3584 CC Utrecht, The Netherlands}
    \affiliation{BASF SE, Carl-Bosch-Strasse 38, 67056 Ludwigshafen am Rhein, Germany}
 \author{Ren\'e van Roij}
 \email{r.vanroij@uu.nl}
	\affiliation{Institute for Theoretical Physics, Utrecht University,  Princetonplein 5, 3584 CC Utrecht, The Netherlands}
 \author{Marjolein Dijkstra}
 \email{m.dijkstra@uu.nl}	\affiliation{Soft Condensed Matter \& Biophysics, Debye Institute for Nanomaterials Science, Utrecht University, Princetonplein 1, 3584 CC Utrecht, The Netherlands}

\date{\today}  

\begin{abstract}
We employ molecular dynamics simulations to investigate ion and water transport driven by an electric field through quasi-two-dimensional nanoslits with a tapered geometry. Despite the absence of surface charge on the (non-polarizable) channel walls and the associated electric double layer, we do observe a robust ion selectivity. This selectivity favors the transport of cations from base to tip when the electric field is directed from base to tip, and anions from base to tip when the field direction is reversed. 
Additionally, we observe a corresponding electro-osmotic water flow from base to tip, regardless of the electric field direction. Intriguingly, ion selectivity and electro-osmotic flow are conventionally associated with surface charge and electric double layers. Here, however, we uncover a novel mechanism for these phenomena in uncharged tapered nanoslits, where ion selectivity arises from the divergence of the heterogeneous water polarization.
\end{abstract}

\maketitle

\section{Introduction}
The past two decades have witnessed a significant miniaturization of microfluidics into the realm of  nanofluidics \cite{Kavokine2021, Bocquet2020}. Ion and water transport  has been studied in nanochannels with various geometries \cite{Siwy20021, Celebi2014, Garaj2010, Hummer2001, Esfandiar2017, Laucirica2020} and surface chemistries \cite{Siwy2004, Stein2004}. Extensive research efforts have focused on voltage-driven electric currents through asymmetric channels with rectifying and diodic properties, such as conical channels with charged channel walls, with dimensions typically ranging from tens to hundreds of nanometers  \cite{Wei1997, Siwy20022, Siwy20021, Siwy2006, Karnik2007, Jubin2018, Boon2022}. The recent development of two-dimensional materials such as graphene \cite{Novoselov2004} has led to the construction of so-called ``nanoslits'', which are quasi two-dimensional channels  with heights on the order  of a single nanometer or even smaller \cite{Keerthi2018, Geim2021}. Water confined to these nanoslits can no longer be realistically viewed as an isotropic dielectric and viscous continuum. Instead, the molecular granularity and the extremely close proximity of (usually weakly polarizable) channel walls are expected to give rise to qualitatively new physical properties and phenomena, that are not captured by structureless continuum theory.  This expectation is confirmed by several recent findings concerning nanoconfined water, such as a reduction of the out-of-plane dielectric constant by more than an order of magnitude \cite{Fumagalli2018}, reduced salt solubility \cite{Robin2021, Zhao2021}, and memristive current hysteresis upon driving by an AC potential \cite{Robin2021}. Interestingly, the memory effect of these quasi-2D channels has been shown to form a basis for circuits exhibiting Hodgkin-Huxley-like voltage spiking \cite{Robin2021} and synapse-like dynamics with Hebbian learning properties \cite{Robin2023}.

\begin{figure}[t!]
    \centering
    \includegraphics[width=\linewidth]{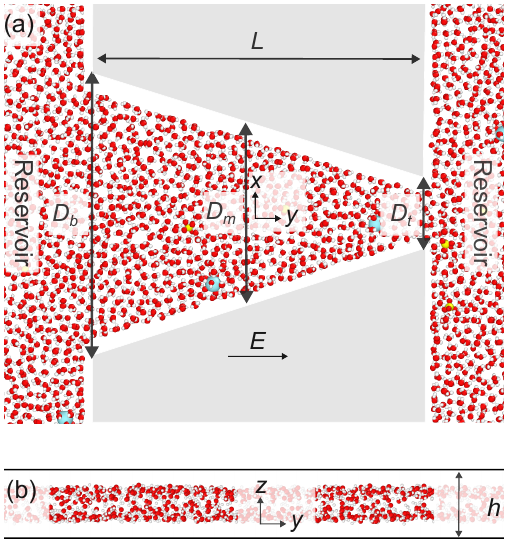}
   \caption{Schematic representation of  (a) a top-down view and (b) a side-view of the quasi two-dimensional geometry of the nanoslit with uniform height $h=1 \text{ nm}$, and length $L=8\text{ nm}$  between tip and base with widths $D_t$ and $D_b$, respectively.  The channel shape is characterized by the mean width $D_m=(D_t+D_b)/2$ and the tip-to-base width ratio $\xi=D_t/D_b$. The channel connects two quasi two-dimensional reservoirs with transport driven by an electric field applied in the $y$-direction. }
  \label{fig:system}
\end{figure}

\begin{figure*}[ht!]
    \centering 
    \includegraphics[]{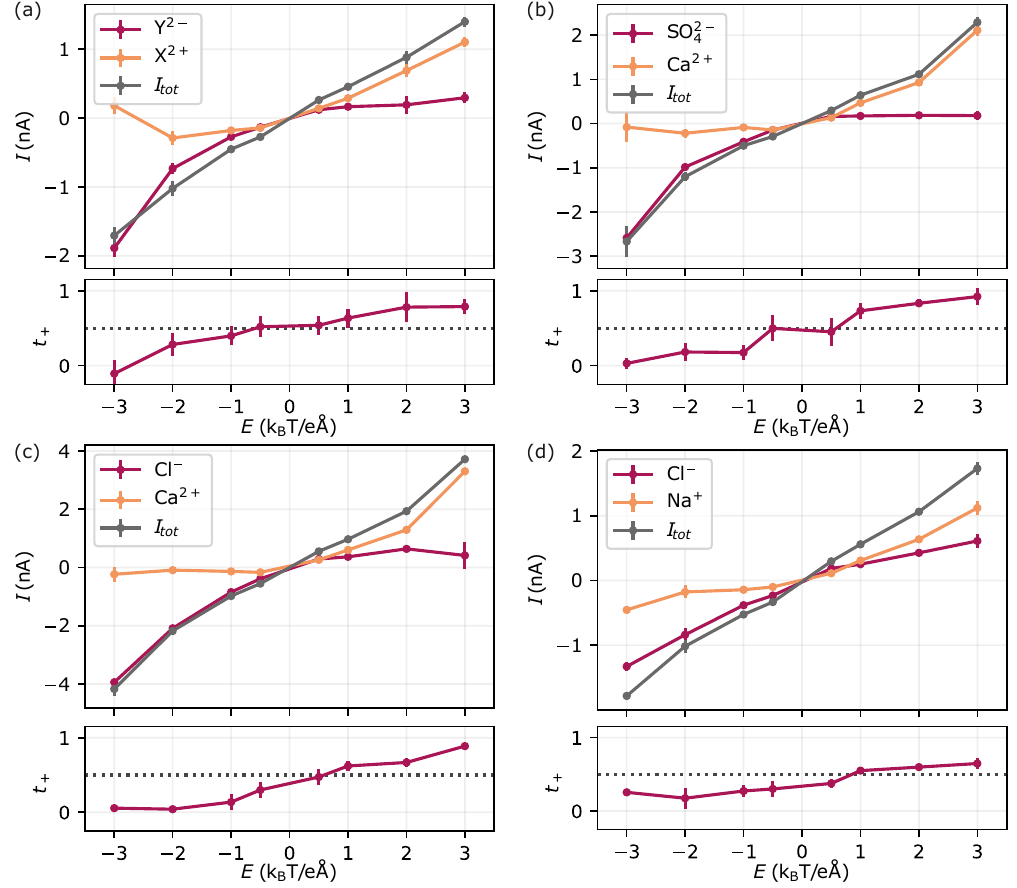}
    \caption{Currents $I$ as a function of the electric field strength $E$ in channels with a tip-to-base width ratio of $\xi=0.25$. The total current $I_{tot}$  (in gray) is further decomposed into the contributions of  negative anions (red) and positive cations (orange). Results are presented for four different salts: (a) a generic divalent salt X$^{2+}$Y$^{2-}$,  (b) calcium sulfate CaSO$_4$, (c) calcium chloride CaCl$_2$, and (d) the monovalent table salt  sodium chloride NaCl. Below each $I-E$ graph, the corresponding transport number $t_+$ is shown. Error bars of the currents represent the standard deviation across five  simulations with different starting configurations. The error bars of the transport number are derived from those of the currents.}
    \label{fig:currents}
\end{figure*}

Despite recent advancements, phenomena such as current rectification and ion selectivity, extensively studied in larger nano- and micro-channels with surface charge on the channel walls, have received limited attention in uncharged quasi two-dimensional systems. Structureless continuum approaches such as the Poisson-Nernst-Planck-Stokes (PNPS) framework do not typically predict these phenomena in  systems without electric double layers. However, the strong confinement to molecular-sized slits fundamentally changes the underlying physics, and new molecular transport mechanisms have been unveiled. For instance, ion selectivity in highly confined geometries has been identified to be caused by entrance effects such as the shedding or rearranging of hydration layers as ions enter the channel \cite{Fornasiero2008, Esfandiar2017, Yu2019, Sahu2017, Xue2022, Li2021, Li2022b, Zhou2023}.

In this work, we present a novel mechanism for selective ion transport through electrically neutral non-polarizable tapered nanochannels, where the selectivity is governed by the polarity of the applied field and stems from the (divergence of the) water polarization. Interestingly, this mechanism also leads to a steady one-way electro-osmotic flow of water, which is transported from base to tip regardless of the polarity of the electric field. We perform  molecular dynamics (MD) simulations to study the flow of ions and water within quasi two-dimensional asymmetric nanoslits with a uniform height $h=1\text{ nm}$. The observed ion selectivity, coupled with a  unidirectional water flow,  allows for enhanced  control over salt and water transport in these channels. We envision that these newly discovered phenomena -- which are not predicted by continuum Poisson-Nernst-Planck-Stokes theory -- could find valuable applications in energy conversion \cite{Siria2017}, water desalination \cite{Werber2016, Li2022b}, and neuromorphic computation \cite{Kamsma2023, Robin2021, Robin2023}.

\begin{figure}[]
    \centering
    \includegraphics[]{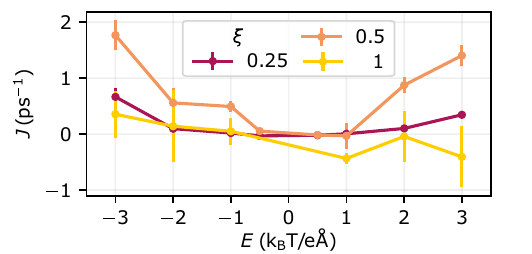}
    \caption{
    The water flux $J$ through channels with the symmetric \xy salt as a function of the electric field $E$ for several tip-to-base ratios $\xi$.}
    \label{fig:transport}
\end{figure}

\section{Quasi-2D tapered channel}
As illustrated in Fig.~\ref{fig:system}, we consider a system consisting of explicit water molecules, cations and anions in two wide reservoirs that are connected by a tapered nanoslit oriented along the $y$-direction that runs perpendicular to the $x$-direction with $x=0$ the symmetry axis. The entire system is confined in the $z$-direction between two planar walls with a fixed height of $h=1 \text{\,nm}$, allowing for only two layers of water molecules \cite{Yoshida2018}. This nanoslit has a fixed length of $L=8\text{\,nm}$ and a width that varies linearly from $D_b$ at the base (located at $y=-L/2$) to $D_t<D_b$ at the tip (positioned at $y=L/2$). We study varying channel geometries, characterized by a  tip-to-base width ratio $\xi=D_t/D_b$, while maintaining the mean width $D_m=\frac{1}{2}(D_b+D_t)$ fixed at $D_m=6\text{\,nm}$. Consequently, the area and the volume of the nanoslit remain constant  throughout this study. The reservoirs have a width of $15 \text{\,nm}$ and a length of $6 \text{\,nm}$, and we apply periodic boundary conditions in the $x-$ and $y-$direction.

\section{Field-driven transport of ions and water}
Using MD simulations, we  investigate  the transport of water and ions driven by an in-plane electric field $E$ applied along the $y$-direction (where  $E>0$ indicates an electric field directed from base to tip) for various channel geometries and dissolved salts. To model the top and bottom walls, as well as the side walls of the channel, we employ a smooth non-polarizable Lennard-Jones 9-3 wall potential between all species (water and ions) and the walls. The water is simulated using  the SPC/E model  \cite{Berendsen2002} at a temperature of $300$ K, while keeping  the areal water density in the system fixed at $\rho_{H_2O} = 24.5$~nm$^{-2}$, which is close to the chemical potential at ambient conditions (see Appendix \ref{AppA}).
Additionally, we consider several types of salts, including  monovalent  NaCl, mono-divalent CaCl$_2$, divalent CaSO$_4$, and a generic  symmetric divalent salt X$^{2+}$Y$^{2-}$. The interactions between ions and between ion and water involve the sum of Lennard-Jones and bare Coulomb potentials, with  all parameters provided in Appendix \ref{AppA}.
The Lennard-Jones parameters for X$^{2+}$ and Y$^{2-}$ are identical to those of calcium, with the exception of the negative sign of the charge of Y$^{2-}$.
We fix the areal density of the cations to $\rho_{X} = 0.07$~nm$^{-2}$ irrespective of the salt type and adjust the areal density of the anions to ensure global charge neutrality. We refer to Appendix \ref{AppA}
for more technical details on the simulations.

\begin{figure}
    \centering
    \includegraphics[width=\linewidth]{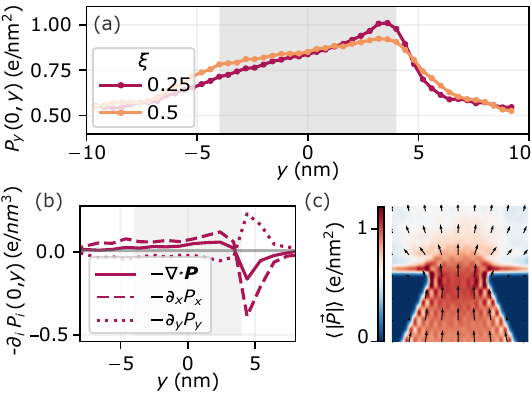}
    \caption{(a) The polarization profile $P_y(0,y)$ along the central axis through the channel for different tip-to-base ratios $\xi$ at an electric field strength of $E=+3$~\ktea. The length of the tapered channel is indicated in gray. (b) Negative divergence of the polarization along the central axis, together with its components $-\partial_x P_x$ and $-\partial_y P_y$. (c) Map of the time-averaged polarization near the tip of the system with a tip-to-base width ratio $\xi=0.25$, at an electric field strength of $E=3$ \ktea. The sidewalls, shown in blue, are non-polarizable.}
    \label{fig:polarization}
\end{figure}

\begin{figure*}[ht!]
    \centering
    \includegraphics[]{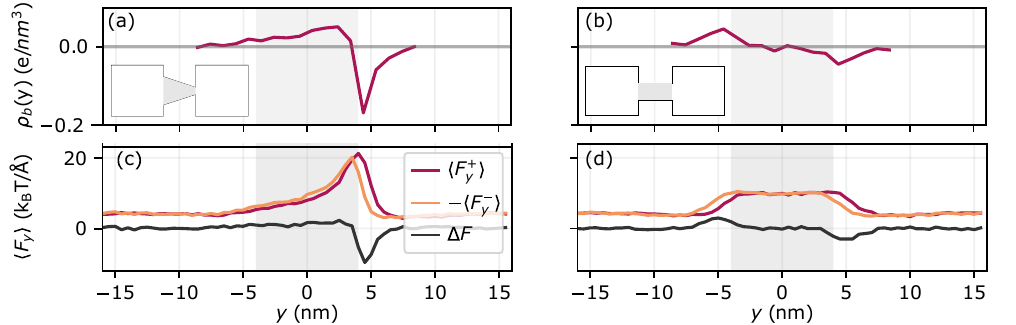}
    \caption{(a-b) The bound charge distributions $\rho_b$  along the central axis of the channel for both  (a) $\xi=0.25$ and (b) $\xi=1$ at an electric field strength of $E = +3$\ktea. (c-d) The forces in the $y$-direction along the channel for both ion species of the divalent salt \xy separately, with their difference $\Delta F \equiv -\langle F_y^-\rangle-\langle F_y^+\rangle$ again for both  (c) $\xi=0.25$ and (d) $\xi=1$ at the same electric field strength, where it should be noted that $\langle F_y^{-}\rangle$ is negative for this value of $E$. Illustrations of the channel geometries are shown in the lower left corners of (a) and (b), and the extent of the channel is shown in gray to illustrate the location of the plotted bound charges and forces.}
    \label{fig:forces}
\end{figure*}

For all four salts under consideration, we first study the transport of ions in a tapered nanoslit with a tip-to-base width ratio $\xi=D_t/D_b=0.25$ as a function of applied field strength $E \in [-3,3]$~\ktea, with $k_{\text{B}}$ the Boltzmann constant and e the elementary charge. This regime corresponds to $E\in[-0.8,0.8] $~V/nm, which is high but experimentally accessible (see Appendix \ref{AppA}). The applied electric field induces ion fluxes from one reservoir through the channel to the other reservoir. For each salt, the total ionic currents through this channel, together with the currents carried by the cations and anions individually, are plotted in Fig. \ref{fig:currents} as a function of electric field strength, averaged over five simulations of $5 \text{~ns}$ each with different starting configurations. To quantify the (a)symmetry in the transport of cations and anions, we also define the cationic transport number $t_+(E) = I_+(E)/I(E)$, where $I(E)$ represents the total ionic current (in the $y$-direction) and $I_+(E)$ represents the current carried by the cations. The transport number for all four salts, as a function of the electric field strength $E$, is also shown in Fig. \ref{fig:currents}.  In a conventional Drude-like picture, where the flux of a given ion species at a given concentration is proportional to the product of its (presumed constant) ion mobility and the strength of the electric field, the transport number $t_+(E)$ of a salt is independent of both the strength and direction of the electric field \cite{Atkins2018}. However, in Fig.~\ref{fig:currents}, where we plot $t_+(E)$ as a function of electric field strength $E$ for various salts, we consistently observe a remarkably strong dependence on the electric field. For negative electric fields, where both $I(E)$ and $I_+(E)$ are negative, we consistently observe that $t_+<0.5$, even reaching $t_+$ smaller than $0.1$ for the divalent cations at $E=-3$\ktea.  As can be seen from the  ion fluxes, this indicates that for negative electric fields ($E<0$), the anions -- moving in the positive $y$-direction --  predominantly contribute to the current, while cations contribute little.  Conversely, for positive electric fields ($E>0$), where both $I(E)$ and $I_+(E)$ are positive, we consistently find $t_+>0.5$ in Fig.~\ref{fig:currents}, with $t_+$ as large as 0.9 for the divalent salts at $E=+3$\ktea. This suggests that for $E>0$, the cations, moving in the direction of the electric field, predominantly determine the current.  Remarkably, this strong ionic selection effect is even present for the generic \xy salt, where both ion species are perfectly symmetric (except for the sign of their charge), such that one might naively expect them to have the same mobility, leading to $t_+=0.5$ for all nonzero $E$ values.

Recent studies have shown a direct link between asymmetries in ionic fluxes, similar to our results in Fig.~\ref{fig:currents}, and the transport of water through tapered channels \cite{Li2021, Zhou2023}. This inspired us to explore the electro-osmotic flux $J$, defined as the average number of water molecules flowing through the channel per unit time. For the generic \xy salt we plot $J$ in Fig.~\ref{fig:transport} as a function of the electric field $E$ for two tapered channels ($\xi=0.25$ and $0.5$) and for a straight channel ($\xi=1$) for comparison. The electro-osmotic fluxes for the other salts are similar and shown in Appendix \ref{AppB}.
Two surprises show up in Fig.~\ref{fig:transport}. The first one is that $J$ is systematically non-zero in the tapered channels, which is unexpected for channels with charge-neutral walls because the conventional mechanism for electro-osmosis relies on the presence of an electric double layer that screens the surface charges \cite{Delgado2005}, see also Appendix \ref{AppC}. The second surprise is that $J>0$ for the tapered channels ($\xi=0.5$ and $0.25$) regardless the polarity of the electric driving field, indicating that the osmotic flow is in all cases directed from base to tip.

\section{water polarization}
This fixed direction of the water flux (from base to tip) regardless of the direction of the electric field is quite unique in non-polarizable channels without surface charge. Similarly, the electric current is  predominantly carried by the ion species moving from base to tip. To the best of our knowledge, this rectification effect has not been observed before. Previously identified mechanisms for ion selectivity in straight channels cannot explain the anomalous transport behavior observed in this tapered system, as discussed in Appendix \ref{AppB}.

Evidently, the electric-field dependent ion selectivity and the associated water flux in this system are induced by the asymmetry in the channel geometry. This is clearly demonstrated in Fig.~\ref{fig:transport}, where there is a marked  difference between the tapered channels ($\xi<1$) and the straight channel ($\xi=1$). By combining the highly non-trivial electrostatics in nanoslits \cite{Fumagalli2018} with the lateral reduction in channel  width, it is conceivable that the height-averaged water polarization ${\bf P}=(P_x,P_y,0)$, where the $z$-component vanishes due to symmetry,  exhibits an anomalous spatial dependence on $x$ and $y$.

Fig.~\ref{fig:polarization}(a) presents the polarization profile $P_y(0,y)$ along the central axis for $\xi=0.25$ and $0.5$, indicating that the polarization near the tip can be about twice as large, and significantly less homogeneous, as in the reservoir, with a more pronounced effect observed for the narrower tip. The heterogeneity of ${\bf P}$ is further analyzed by studying the $y$-dependence of $\partial P_x/\partial x$ and $\partial P_y/\partial y$ along the central axis in Fig.~\ref{fig:polarization}(b). This analysis reveals not only their opposite signs but also the dominance of  $\partial P_x/\partial x$, indicating that the height-averaged bound charge distribution $\rho_b(x,y)\equiv - \nabla\cdot{\bf P}(x,y)$ is nonzero \cite{Griffiths2017}, see also Appendix \ref{AppD}.  This polarization distribution is further visualized in Fig. \ref{fig:polarization}(c), where the vector field ${\bf P}(x,y)$  is plotted near the tip of the channel, over the heat map of $|{\bf P}(x,y)|$ (colors) for the salt-free system at $\xi=0.25$ and $E=+3$\ktea.

\begin{figure*}[t!]
    \centering
    \includegraphics[width=\linewidth]{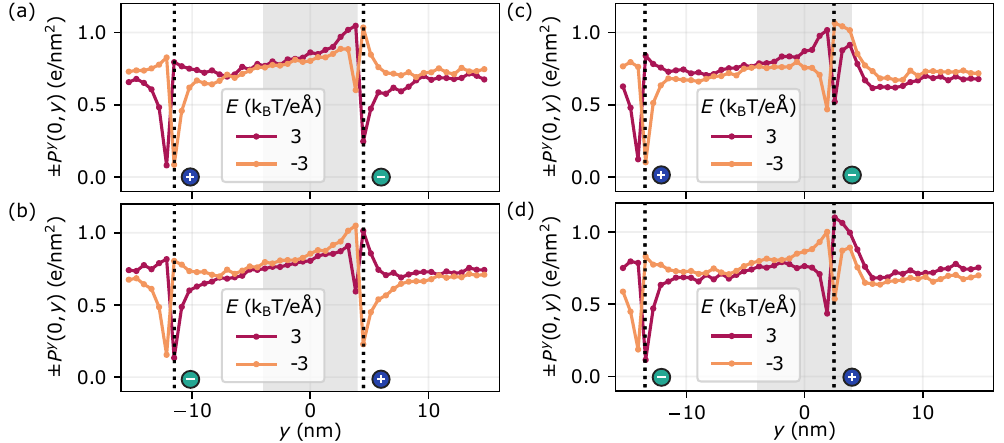}
    \caption{The time-averaged polarization component $P_y$ along the central axis of a channel with $\xi=0.25$, for two electric field strengths $E$ (red and yellow curves), in the presence of a pair of Y$^{2-}$ and X$^{2+}$ ions on the central axis at static $y$-coordinates $y_-$ and $y_+$, respectively,  given by (a) $y_-= 4.5\text{~nm}$ and $y_+=-11.5 \text{~nm}$, (b) $y_-=-11.5\text{~nm}$ and $y_+=4.5 \text{~nm}$, (c) $y_-=2.5\text{~nm}$ and $y_+=-13.5\text{~nm}$, and (d) $y_-=-13.5\text{~nm}$ and $y_+=2.5\text{~nm}$. Here $y_\pm$ is indicated by the dashed line, the sign of the charge of the ion is illustrated with circles containing a plus or minus. The extent of the channel is shown in gray.}
    \label{fig:perturbed_polarization}
\end{figure*}

In Figs. \ref{fig:forces}(a) and (b), we plot $\rho_b(0,y)$ along the central axis of the channel for $E = +3$\ktea, in (a) for the tapered channel with $\xi=0.25$ and in (b) for the straight channel with $\xi=1$.  In (a) we observe a significant negative bound charge density just outside the tip of the channel, and inside the channel a much lower and a spatially much more extended positive charge density that extends into the left reservoir.  In contrast, the straight channel in (b) shows an equal but opposite bound charge distribution around the inlet and outlet. The (divergence of the) heterogeneous polarization profile may impact ions and their hydration layers. To quantify the impact of this water-polarization effect, we use constrained MD simulations of a single pair of dissolved ions to calculate the thermal average $\langle F_y^\pm\rangle(y)$ of the $y$-component of the water- and field-induced force that acts on a  cation ($+$) or anion ($-$) at position $y$ on the central axis of the channel. We focus on the generic symmetric salt \xy and keep the ionic pair sufficiently far apart to minimize their direct interaction while keeping the system globally charge-neutral (see Appendix \ref{AppE}). For $E=+3$~\ktea, Figs. \ref{fig:forces}(c) and (d) show the resulting force curves $\pm\langle F_y^\pm\rangle(y)$, in (c) for the tapered channel ($\xi=0.25$) and in (d) for the straight one ($\xi=1$). As expected,  the forces on the cations and anions have opposite signs and a larger magnitude inside both channels than in the reservoirs due to field-line compression in both channels. However, in the tapered channel the enhancement factor of the forces is much larger and, importantly, develops an asymmetry not only between cations and anions but also between tip and base.  Specifically, at $E=+3$~\ktea~ the outward force on the cations at the tip of the tapered channel is much larger than the inward force on the anions at the tip, whereas the force on the cations well inside the channel is only marginally smaller than on the anions. This polarization-induced force asymmetry between the tip and base is key to the observed selectivity, as it favors the transport of cations over anions for the present case $E=+3\text{\ktea}>0$ in a tapered channel as observed in Fig.~\ref{fig:transport}(a). Similarly,  anion transport is favored at $E<0$. This combined tip-base and anion-cation asymmetry disappears in the straight channel, as shown in Fig.~\ref{fig:forces}(d). Here the excess positive force on the cations at $y=L/2$ is compensated by an equally large excess negative force on the anions at $y=-L/2$ such that both ionic species contribute equally to the electric current (i.e.  $t_+\approx 0.5$).

We quantify the force asymmetry by the difference $\Delta F\equiv -\langle F_y^-\rangle-\langle F_y^+\rangle$, which is also plotted in Fig.~\ref{fig:forces}(c) and (d) as a function of $y$. Interestingly, $\Delta F(y)$ shows a remarkable similarity to the bound charge density profile $\rho_b(y)$ of the salt-free systems shown in Fig.~\ref{fig:forces}(a) and (b), respectively,  for both the tapered and straight channel. The direct connection between $\Delta F(y)$ and $\rho_b(y)$ is also evident when we realize that $\Delta F$ would be identically zero for the perfectly symmetric salt \xy in a structureless dielectric continuum, where the magnitude of the Coulomb force is independent of the sign of the ionic charge. In the explicit SPC/E water model we consider here, its nontrivial (divergence of the) polarization profile in the asymmetric tapered channel couples differently to the cationic compared to the anionic hydration layers. 

This is shown explicitly in Fig. \ref{fig:perturbed_polarization} where we plot the water polarization along the channel at an electric field strength of $E=\pm3$~\ktea, in the presence of a fixed pair of ions on the central axis. In Fig. \ref{fig:perturbed_polarization}(a), the anion is fixed at $y_-=4.5$~nm and the cation at $y_+=-11.5$~nm, such that the anion is located where $\nabla \cdot {\bf P}$ is largest while the cation is in the reservoir where $\nabla \cdot {\bf P} = 0$. We see that the polarization around the anion is much less perturbed for $E<0$ than for $E>0$, whereas the difference between the perturbations vanishes around the cation in the reservoir. The reverse holds true upon exchanging the cation and the anion position, as is shown in Fig. \ref{fig:perturbed_polarization}(b), where the polarization profile with a cation at $y_+=4.5$~nm gets perturbed much less for $E>0$ compared to $E<0$. In Fig. \ref{fig:perturbed_polarization}(c-d) we see an opposite but much smaller effect when one of the ions is placed right inside the channel, where the divergence of the polarization $\nabla\cdot \bf P$ has the opposite sign. When the cation is placed at $y_+=2.5$~nm, we see that the polarization is slightly less perturbed for $E<0$ compared to $E>0$, and conversely, when an anion is placed at $y_-=2.5$~nm, the polarization is slightly less perturbed for $E>0$. The difference in disturbance of the polarization profile due to the ions appears to be proportional to the divergence of the polarization in the channel. Therefore, we can directly attribute the force acting on the ions to the degree to which the ions distort the polarization profile, which is asymmetric with respect to the ion charge when there is a non-zero divergence, leading to ion selectivity.

\section{Conclusions}
In conclusion, our molecular dynamics simulations of transport of aqueous electrolyte films through a tapered nanochannel exhibit strong ion selectivity, despite the absence of any surface charge on the channel walls. The favored ion is determined by the direction of the electric field, hence making the ion selectivity tunable. Interestingly, the ion species traveling from the base to the tip is always favored, resulting in a water flux aligned in that direction, irrespective of the electric field direction. To the best of our knowledge, this remarkable phenomenon has not  been observed before in non-polarizable channels, nor have they been predicted theoretically. Advances in the theory of structured dielectrics \cite{Maggs2006, Paillusson2010, Berthoumieux2019, Blossey2022} could potentially be applied to the system considered in this work. We show that the average force exerted on a cation in a tapered channel differs significantly from that on an anion due to the divergence of the water polarization. This discrepancy arises from the distinct polarization profiles at the tip and base of the tapered channel, driven by their different widths. Ultimately, these differences in the forces acting on cations and anions lead to the observed ion selectivity and the associated water fluxes. This novel mechanism could find applications in diverse fields such as water desalination \cite{Werber2016}, energy conversion \cite{Siria2017}, neuromorphic computation \cite{Robin2021}, and in the further development of nanofluidic devices  \cite{Robin2023}.

Further research into tapered quasi-2D nanoslits could investigate the dependence of the transport properties and polarization heterogeneity on the dimensions of the system.  Smaller channel dimensions may induce steric effects on the water polarization profiles and hinder the alignment of the water molecule dipoles as discussed in this work. Furthermore, a full comparison of relevant quantities such as the dielectric permittivity, thermodynamics, structure, and transport properties of various water models and simulation force fields to those found in experiments in quasi-2D slits is currently lacking in the literature and would be an interesting future direction of research.

\section*{Acknowledgements}
T.E.V. and M.D. acknowledge funding from the European Research Council (ERC) under the European Union's Horizon 2020 research and innovation programme (Grant agreement No. ERC-2019-ADG 884902 SoftML). G. C.-V. acknowledges funding from The Netherlands  Organization  for  Scientific  Research  (NWO) for the ENW PPS Fund 2018 - Technology Area Soft Advanced Materials ENPPS.TA.018.002.  During the execution of the project, G. Giunta received funding from The Netherlands Center for Multiscale Catalytic Energy Conversion (MCEC), an NWO Gravitation program funded by the Ministry of Education, Culture and Science of the Government of The Netherlands.

\section{Author declarations}

\subsection{Conflicts of Interest}
The authors have no conflicts to disclose.

\subsection{Author Contributions}
\textbf{Tim E. Veenstra}: Conceptualization (equal), Data curation (equal), Formal Analysis (equal), Investigation (equal), Writing - original draft (equal), Writing - review and editing (equal); \textbf{Gerardo Campos-Villalobos}: Conceptualization (equal), Writing - original draft (equal), Writing - review and editing (equal); \textbf{Giuliana Giunta}: Conceptualization (equal), Writing - original draft (equal), Writing - review and editing (equal); \textbf{Ren\'e van Roij}: Conceptualization (equal), Writing - original draft (equal), Writing - review and editing (equal); \textbf{Marjolein Dijkstra}: Conceptualization (equal), Funding acquisition (equal), Writing - original draft (equal), Writing - review and editing  (equal).

\subsection{Data Availability}
The data that support the findings of this study are available from the corresponding author upon reasonable request.




\appendix

\setcounter{secnumdepth}{4}

\renewcommand{\theequation}{A\arabic{equation}}
\begin{center}
\begin{table*}[t!]
    \centering
    \begin{tabular}{ p{3cm} p{2cm} p{2cm} p{2cm} p{2cm}} \vspace{5pt}
      & $\varepsilon$ (kcal/mol) & $\sigma$ (\angstrom) & $q/e$ & $m$ (g/mol) \vspace{5pt} \\ \hline 
     O (H$_2$O) & 0.1553 & 3.166 & -- 0.8476 & 15.9994 \\  
     H & 0 & 1 & + 0.4238 & 1.008 \\
     Na & 0.123 & 2.35 & + 1 & 28.990 \\
      Cl & 0.1 & 4.401 & -- 1 & 35.453 \\
     S & 0.25 & 3.55 & + 2 & 32.065 \\
     O (SO$_4$) & 0.1553 & 3.166 & -- 1 & 15.9994 \\ 
     Ca & 0.1 & 2.895 & + 2 & 40.078 \\
      X & 0.1 & 2.895 & + 2 & 40.078 \\
     Y & 0.1 & 2.895 & -- 2 & 40.078 \\
     LJ 9-3 wall & 0.478 & 3.214 & 0 & $\infty$ \\

    \end{tabular}
    \vspace{10pt}
    \caption{Lennard-Jones parameters, charge, and mass of all atoms  used in the simulations.}
    \label{tab:LJ}
    \end{table*}
\end{center}

\section{Simulation details}
\label{AppA}
We perform molecular dynamics simulations using the Large-scale Atomic/Molecular Massively Parallel Simulator (LAMMPS) software \cite{Thompson2022}.  In our simulations, we describe the atom-atom  interactions  and the interactions between atoms and  channel  walls using Lennard-Jones potentials. 
For the atom-atom interactions, we use the  standard Lennard-Jones 12-6 interaction potential with parameters  taken from  Robin \textit{et al.}  \cite{Robin2021}. When considering the interactions of an atom with the channel walls, we employ a Lennard-Jones 9-3 potential. 
This Lennard-Jones 9-3 potential is derived by integrating the Lennard-Jones 12-6 interactions of the considered atom with all the atoms in an (effectively infinite) wall of Lennard-Jones particles, characterized by the density $\rho$, effective interaction energy $\varepsilon_{eff}$,  and length scale $\sigma$. This wall interaction potential is given by
\begin{equation}
    \Phi_{wall}(d) = \varepsilon_{eff} \left[ \frac{2}{15} \left( \frac{\sigma}{d} \right)^9 - \left(\frac{\sigma}{d}\right)^{3} \right],
\end{equation}
where $d$ is the distance measured perpendicular to the wall. The effective interaction energy $\varepsilon_{eff}$ for the Lennard-Jones 9-3 wall potential corresponding to the smoothed out graphite walls, is calculated with 
\begin{equation}
    \varepsilon_{eff} = \frac{2\pi}{3}\rho_C \sigma_C^3 \varepsilon_C,
\end{equation}
where the average carbon density in graphite is $\rho_C~=~0.121$~\angstrom$^{-3}$, computed using the interlayer distance and  bond length of graphite \cite{Xu2013}. The other parameters of carbon are   taken from Ref.~\cite{Robin2021}: $\sigma_C = 3.214$ \angstrom~ and $\varepsilon_C = 0.0566$ kcal/mol. The mass of the wall is assumed to be infinite. Effectively, this means that we simulate the walls as an infinitely large coarse-grained slab of graphite, inspired by recent work on graphene and graphite nanochannels \cite{Keerthi2018, Geim2021}, although the specifics of the confining atom-wall interaction potential have only a minor influence on the liquid properties \cite{Kumar2007}. Charged particles interact also via the Coulomb potential. We explicitly modeled the water molecules using the SPC/E model \cite{Berendsen2002}. The parameters for all these interactions are shown in Table  \ref{tab:LJ}. Importantly, we consider the wall to be non-polarizable such that no charge will accumulate on the surfaces when electric fields are applied.

The default Good-Hope geometric mixing rule was used to calculate the interactions between different atoms, such that the effective length and energy scales for the interaction between atoms of type $i$ and $j$ are given by $\sigma_{ij} = \sqrt{\sigma_i\sigma_j}$ and $\varepsilon_{ij} = \sqrt{\varepsilon_i\varepsilon_j}$ \cite{Good1970}. We also simulate two covalently bonded molecules: water (H$_2$O) and sulfate (SO$_4$). The water molecules are kept rigid using the SHAKE algorithm, while the sulfate molecules are kept rigid using the LAMMPS `rigid' command,  as SHAKE cannot handle molecules that consist of more than four atoms. Both molecules have tetrahedral internal angles of $\theta = 109.47^\text{o}$. Water has a bond length of $r_{OH} = 1.00$~\angstrom, and sulfate has a bond length of $r_{SO} = 1.49$~\angstrom. A constant temperature of $300 \text{\,K}$ was maintained using  the Nos\'e-Hoover thermostat. A cutoff of $1\text{\,nm}$ was applied for the short-ranged interactions, while long-ranged interactions were computed using the Particle-Particle Particle-Mesh (PPPM) method. The areal water density $\rho_{H_2O}$ in the nanoslit with Lennard-Jones 9-3 walls was determined using grand-canonical Monte Carlo simulations in LAMMPS at a chemical potential of $\mu=-10.4 \text{\,kcal/mol}$ \cite{Whitley2004}, which was shown to reproduce bulk densities of SPC/E water. This resulted in an areal density of $24.5 \text{\,nm}^{-2}$. The  chemical potential was determined  for bulk water at a temperature of $298 \text{\,K}$. 
Simulations were first run for $1 \text{~ns}$, which we verified to be sufficient to reach steady state, after which the relevant quantities were measured for $4 \text{~ns}$. To improve statistics, the simulations used to obtain the transport properties, shown in, among others, Figs. \ref{fig:currents} and \ref{fig:transport} were repeated 5 times. The errors in the currents and fluxes are determined from the standard deviation of the measurements across 5 independent simulations, and the error in the transport number $t_+$ is obtained from calculating the error propagation from those of $I_+$ and $I_-$.

In this study, we investigate the transport of ions and water driven by an in-plane electric field $E$ for various channel geometries and dissolved salts. 
The magnitude of the electric field strengths that are applied are in the range $|E| \in (0.25, 3)$ \ktea, which at room temperature corresponds to $|E| \in (0.065, 0.78) \text{V/nm}$. Although these field strengths are relatively high, electric fields on the same order of magnitude, up to $0.57 \text{~V/nm}$, have been realized in experimental systems, and therefore should be accessible. In Tab. \ref{tab:experiments}, we list several experimental studies and the maximum electric field strengths that have been achieved. In our study, as in several other simulation studies \cite{Li2021, Zhou2023, Yu2019}, we use this relatively high range of field strengths in order to reduce the computational costs to obtain reasonable signal-to-noise statistics for the electric current, which is challenging given the relatively large number of water molecules per ion.

\begin{table}[]
    \centering
    \vspace{20pt}
    \begin{tabular}{l|c}
      Study & Strength $E$-field (V/nm) \vspace{1pt} \\ \hline          Fragasso \textit{et al.} \cite{Fragasso2019} & $0.2$ \\
        Heerema \textit{et al.} \cite{Heerema2015} & $0.13 - 0.44$\\
        Chuang \textit{et al.} \cite{Chuang2024} & $0.57$ \\
        
    \end{tabular}
    \caption{Several experimental studies and the maximum field strengths achieved in these systems.}
    \label{tab:experiments}
\end{table}

\renewcommand{\theequation}{B\arabic{equation}}
\section{Currents, Transport Numbers, and Fluxes}
\label{AppB}
As described in the main text and shown in Fig.~2 of the main text, we observe anomalous behavior in the ion fluxes and the transport number $t_+(E) = I_+(E)/I(E)$ as a function of the ($y$-component of the) external electric field strength $E$,  where
$I(E)=\langle \sum_i q_i e v_{y,i}\rangle/L_y$ represents the total ionic current (in the $y$-direction) with $q_i$ and $v_{y,i}$  the charge and velocity in the $y$-direction of ion $i$, respectively, $L_y$ the length of the simulation box in the $y$-direction, and $I_+(E)$ represents the current carried by the cations. For electric fields pointing from base to tip, $E>0$, most of the current is carried by the positive cations moving from base to tip, while for electric fields pointing from tip to base, $E<0$,  the current is predominantly carried by the negative anions moving also from base to tip. For the generic divalent salt, the transport number even reaches negative values at $E=-3$~\ktea. At these salt concentrations of a divalent salt, there is some mild ion clustering, leading to the formation of effectively neutral ion pairs. In this particular case, the water flow is sufficiently strong, and the ion selectivity  sufficiently large, that more cations bound in neutral pairs are advected in the positive $y$-direction by the fluid flow than unbound cations are driven  in the negative $y$-direction by the electric field. 

For all four salt species we see a strong selectivity, with cations moving towards the tip dominating the (positive) current at $E>0$, and anions moving towards the tip dominating the (negative) current at $E<0$. The effect is weakest, but still significant,  for the monovalent salt NaCl. This phenomenon is unique to these tapered channels. 

\begin{figure*}[t!]
    \centering    \includegraphics{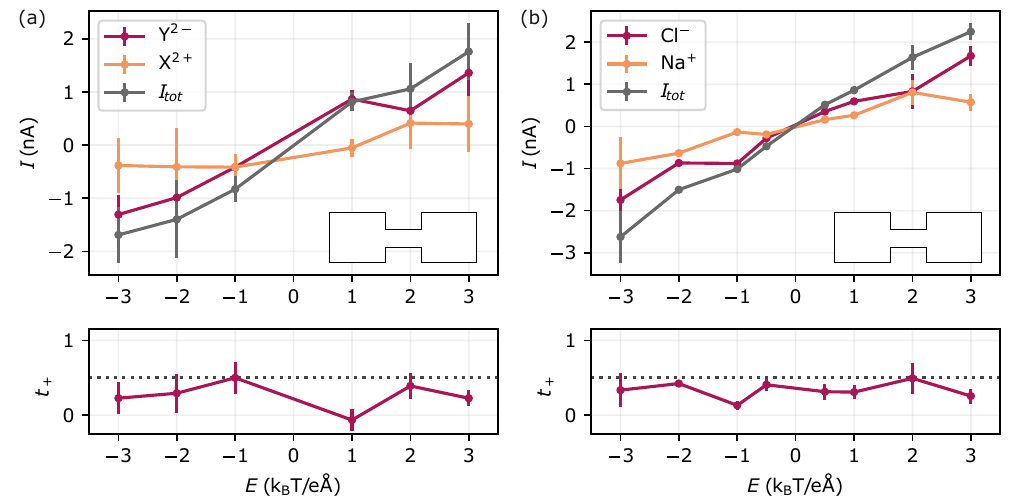}
    \caption{Currents $I$ as a function of the electric field strength $E$ in channels with a tip-to-base width ratio of $\xi=0.25$. The total current $I_{tot}$  (in grey) is further divided  into the contributions of the negative anions (red) and positive cations (orange). Results are presented for both a divalent and a monovalent salt: (a) a generic divalent salt \xy, and (b) monovalent sodium chloride NaCl. Below each $I-E$ graph, the corresponding transport number $t_+$ is shown.}
    \label{fig:currents_SI_xi1}
\end{figure*}

For the generic divalent salt \xy and for the monovalent salt NaCl, we also performed these simulations in straight channels with a tip-to-base ratio of $\xi=1$ and a rectangular cross section of $1\times6$ nm$^2$. The results are shown in Fig.~\ref{fig:currents_SI_xi1},
where we observe that the transport number remains approximately constant as a function of the electric field strength, as  expected. Cations seem to have a lower mobility than anions in these channels, as their contributions to the total current consistently lag behind  that of the anions, as reflected by $t_+$ consistently staying  below $0.5$. Recent work has shown, however, that ion selectivity in  cylindrical channels strongly depends on the channel radius  \cite{Zhou2023}. It is therefore conceivable that straight channels with different heights and widths may exhibit qualitatively different ion selectivity properties. 

\begin{figure}
    \centering   
    \includegraphics{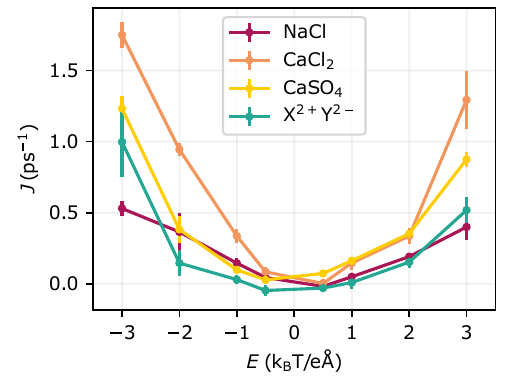}
    \caption{Water flux $J$ as a function of electric field strength $E$ for all four salts considered here, in a channel with a tip-to-base ratio $\xi=0.25$.}    \label{fig:water_flux_all_salt}
\end{figure}

In Fig.~3 of the main text, we show the water flux $J$ in a system containing the generic divalent salt \xy. We observed an electro-osmotic flow  directed from base to tip, so $J>0$ irrespective of the direction of the external electric field. This behavior is not specific to the \xy salt. In Fig.~\ref{fig:water_flux_all_salt}, we show the water fluxes in systems containing all four different salts, and we consistently observe that $J>0$. This demonstrates that the uni-directional flow is robust and insensitive to ion charge, as it occurs with both monovalent, mono-divalent and divalent salts. It should be noted here that the water flux is roughly two orders of magnitude larger than the relative ionic current $I_{cat} - I_{an}$, and therefore cannot just be explained by the transport of ions that carry a hydration shell.

\begin{figure}
    \centering
    \includegraphics{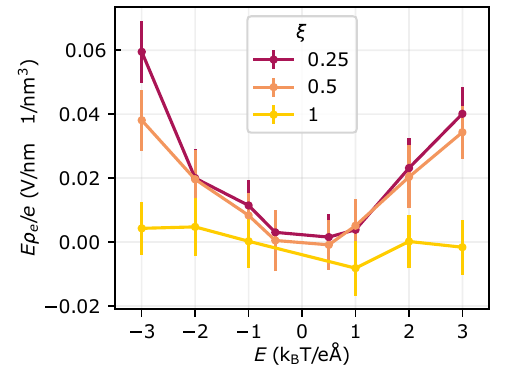}
    \caption{The product of the ionic charge density $\overline{\rho_e}$ and the electric field $E$, for the perfectly symmetric \xy salt as a function of the electric field $E$ for several tip-to-base ratios $\xi$.}
    \label{fig:IonCharge_SI}
\end{figure}

Because a nonzero electro-osmotic flux requires an electric body force on mobile charges, and hence a non-zero ionic charge distribution $\rho_e(x,y,z)$ in the channel (i.e. within the interval $y\in[-L/2,L/2]$), we measure its volumetric average $\overline{\rho_e}$. This is calculated using the time-averaged number of cations and anions in the range $y \in (-L/2, L/2)$. Although one might expect $\overline{\rho_e}$ to be (essentially) zero due to  the charge-neutral walls and the absence of electric double layers, the plot of the (scaled and channel-averaged) body force $E\overline{\rho_e}$ as a function of $E$ shown in Fig.~\ref{fig:IonCharge_SI} is clearly nonzero for the tapered channels ($\xi=0.5$ and $0.25$). In fact we also observe that the volume-averaged electric body force  is positive, $E\overline{\rho_e}>0$,  regardless of the polarity of the electric field, which clearly aligns with our observation that $J>0$. Furthermore,  this positive average body force implies that the net ionic charge in the channel is positive for $E>0$ and negative for $E<0$, which agrees, at least qualitatively, with the ion selectivity that followed from the dependence of the transport number $t_+$ on $E$ shown in Fig.~2 of the main text. This indicates that the current is predominantly carried by the most abundant ionic species in the channel, assuming equal intrinsic mobilities of \xy ions. 
Interestingly, this also agrees with the findings of Zhou \textit{et al.}, who showed that ion selectivity in channels without a surface charge can lead to non-zero net ion charge densities \cite{Zhou2023}.

\begin{figure}[t]
    \centering
    \includegraphics{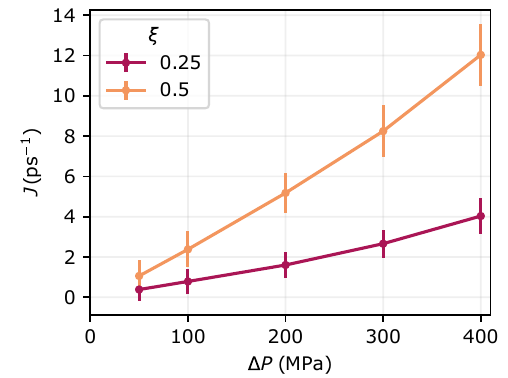}
    \caption{Water flux $J$ as a function of the applied pressure difference $\Delta P$ for  channels with tip-to-base ratios of $\xi=0.25$ and $\xi=0.5$.}
    \label{fig:dP}
\end{figure}

In Fig.~\ref{fig:IonCharge_SI} we also observe that the electro-osmotic driving force $E\overline{\rho_e}$ is similar in magnitude for both tip-to-base ratios $\xi=0.25$ and $\xi=0.5$. At first sight this seems to be inconsistent with the much higher water flux in the less tapered channel with $\xi=0.5$ compared to $\xi=0.25$, as shown in Fig.~3 of the main text. Here, however, we show that this apparent discrepancy arises from the significantly higher friction in the more tapered channel with $\xi=0.25$. We perform simulations in the $NVT$ ensemble in the water-only system without any ions and at zero electric field, but instead apply a pressure difference across the channel. Following the  methodology of Refs.~\cite{Zhu2002, Zhou2023}, this is achieved by exerting a force $f$ on the oxygen atoms of water molecules within a $4$~nm zone in one of the reservoirs, far from the channel. This method does not apply a force to each water molecule in the system, as that could unphysically influence the trajectories of particles within the channel and alter the flow profile. Instead a larger force is exerted on a small portion of the molecules, far from the channel, to obtain the same pressure difference: 
\begin{equation}
    \Delta P = \frac{n f}{A_{eff}}, 
\end{equation}
where $n$ represents the number of molecules in the $4$~nm zone and $A_{eff}$ the effective cross-sectional area of the nanoslit. We use an effective channel height of $5$  \AA, determined by considering the maximum and minimum coordinates  of all oxygen atoms in a similar simulation. This results in an effective cross-sectional area of approximately $A_{eff} \approx 150\times 5 = 750$ \AA$^2$. We apply a pressure difference and allow the system to reach a steady state over a period of  $50$ ps. Subsequently, we run the simulation for $5$ ns during which we measure the water flux by counting the water molecules that pass the $y=0$ center-line of the channel. We applied pressures ranging from $50$ to $400$ MPa, both for $\xi=0.25$ and $\xi=0.5$, and the resulting water fluxes are shown in Fig.~\ref{fig:dP}.  We clearly observe that the water flux at the same driving pressure drop $\Delta P$ is approximately $3$ times higher for the channel with a tip-to-base width ratio of $\xi=0.5$ compared to the more strongly tapered channel with $\xi=0.25$. 
The fact that the water friction seems to be roughly $3$ times higher in the more tapered channel aligns well with the observation in Fig.~3 of the main text that the $\xi=0.5$ channel exhibits a significantly higher water flux despite having a comparable or slightly lower ionic charge density within the channel.

\begin{figure}
    \centering
    \includegraphics{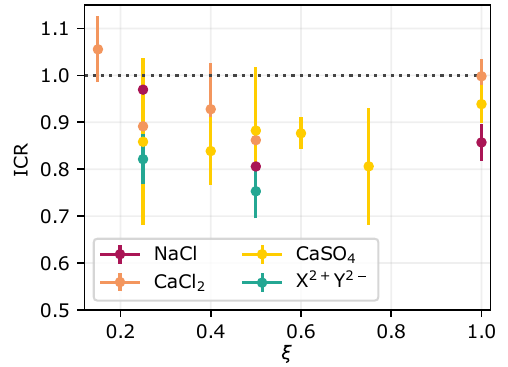}
    \caption{The current rectification ratio $ICR$ (as defined in the text) as a function of tip-to-base ratio $\xi$ and for various salts.}
    \label{fig:ICR_SI}
\end{figure}

In Fig.~\ref{fig:IonCharge_SI}, we note a (slightly) larger body force at $E=-3$\ktea \, compared to $E=+3$\ktea. Here we show that a similar inequality holds true for the ionic current. We plot the ionic current rectification $\text{ICR}~=~|I(E=3$~\ktea$)/I(E=-3$~\ktea$)|$ as a function of $\xi$ for all four salts that we consider in Fig.~\ref{fig:ICR_SI}. For the entire range of different tapered geometries and for all salts we consistently observe that $\text{ICR}<1$, which implies that the channel exhibits a higher conductivity when the electric field points from tip to base. This is fully consistent with the higher electro-osmotic force in the channel for $E<0$ compared to $E>0$, as shown in Fig.~\ref{fig:IonCharge_SI}.

\section{Structureless continuum calculations}
\label{AppC}
\begin{figure}
    \centering
    \includegraphics[width=\linewidth]{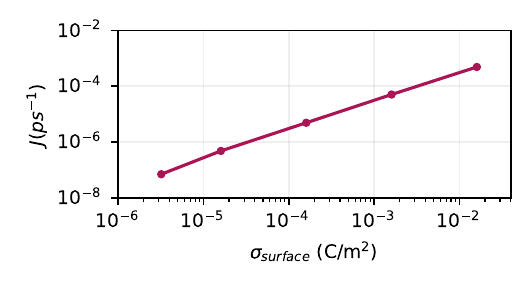}
    \caption{Water flux $J$ calculated with the Poisson-Nernst-Planck-Stokes equations, as a function of surface charge $\sigma_{surface}$, at an electric field strength of $E = 0.2585$~V/nm.}
    \label{fig:continuous-fluid}
\end{figure}

 To highlight the novelty of our findings, we can compare these results to those found in structureless continuum theory. To this end, we compare the fluid flow and the cationic transport number to those in a Poisson-Nernst-Planck-Stokes framework, as calculated with COMSOL Multiphysics \cite{COMSOL}. We use the tapered slit geometry with a tip-to-base width ratio $\xi=0.25$ and apply an electric field $E = 0.2585$~V/nm. The ionic diffusion coefficient is taken to be $D=10^{-9}$~m$^2$/s for all ionic species. At these high electric field strengths, the calculations for a strictly zero surface charge turn out to be unstable and did not converge. Therefore we consider a decreasing surface charge ramp that ranges all the way from $1.6 \cdot 10^{-2}$~C/m$^2$ down to $3.2 \cdot 10^{-6}$~C/m$^2$, where the surface charge is only applied to the sides of the tapered channel, not to the top or bottom surfaces. We calculate the water flux based on the calculated velocity profiles using the same water density used in the MD simulations of $25.4$~nm$^{-2}$. The resulting fluid fluxes as a function of surface charge are shown in a log-log plot in figure \ref{fig:continuous-fluid}, from which we can see that the water flux decreases linearly with decreasing surface charge. From this we can conclude that the expected water flux in the Poisson-Nernst-Planck-Stokes framework is indeed zero in the case of zero surface charge. Furthermore, for the surface charges used in these calculations, the water flux is already orders of magnitude smaller than those found in MD simulations without any surface charge, as can be seen by comparing Fig.~\ref{fig:continuous-fluid} with Fig.~\ref{fig:water_flux_all_salt}.

\begin{figure}
    \centering
    \includegraphics[width=\linewidth]{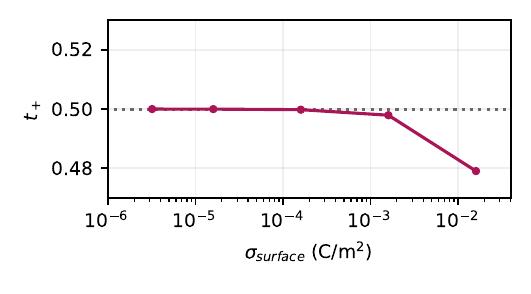}
    \caption{The cationic transport number $t_+$ calculated with the Poisson-Nernst-Planck-Stokes equations as a function of surface charge $\sigma_{surface}$ at an electric field strength of $E = 0.2585$~V/nm.}
    \label{fig:continuity-transport}
\end{figure}

Furthermore, we look at the cationic transport number in this system and plot it for the same range of surface charges in Fig. \ref{fig:continuity-transport}. We see at the highest surface charge an imbalance in the cationic and anionic fluxes, as the transport number is (slightly) smaller than $0.5$. This is due to the negative ionic charge in the electric double layer that generates a negative electro-osmotic current for $E>0$. This effect disappears when the surface charge is lowered, leading to a transport number of $0.5$. On the basis of  these continuum calculations we conclude that a tapered channel without any surface charge is not expected to exhibit ion selectivity. The electrokinetic phenomena observed in our MD simulations of explicit water must therefore be driven by an electro-osmotic flow mechanism that is completely different from the conventional electric body force on the fluid generated by the electric double layer.\\

\label{hydration}

\begin{figure}[t!]
    \centering
    \includegraphics{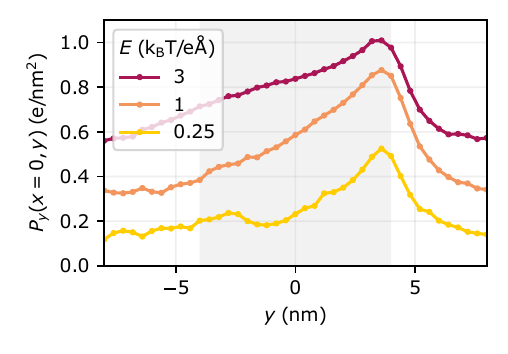}
    \caption{The $y$-component $P_y$ of the polarization of water along the central $y$-axis of a channel with tip-to-base ratio $\xi=0.25$ at three different electric field strengths. The shaded area indicates the extent of the channel.}
    \label{fig:polarization-SI}
\end{figure}

\begin{figure}[t!]
    \centering
    \includegraphics{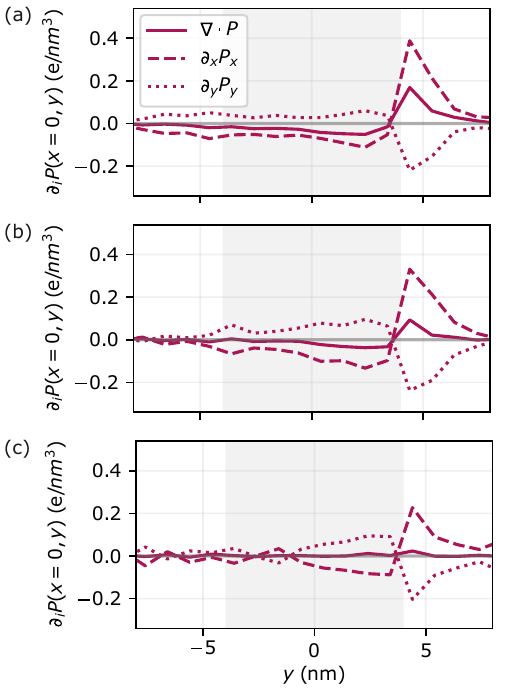}
    \caption{The derivatives of the polarization of water along the central $y$-axis of a channel in a salt-free system with a tip-to-base ratio $\xi=0.25$ at electric field strengths (a) $3$ \ktea, (b) $1$ \ktea~ and (c) $0.25$ \ktea. The shaded area indicates the extent of the channel.}
    \label{fig:divergence-polarization-SI}
\end{figure}

\section{Water Polarization}
\label{AppD}
When external electric fields are applied, water molecules polarize due to the tendency to align their permanent dipole moment with the field. However, the electrostatic behavior in our system is highly non-trivial, particularly due to the combination of strong confinement in the $z$ direction and the tapered geometry of the channel. Recent experimental work has demonstrated that water confined within nanoslits can exhibit an anomalously low and anisotropic dielectric constant \cite{Fumagalli2018}. The decreasing width of the channels in our system, combined with anisotropic permittivity, leads to anomalous behavior. We study these effects by performing simulations in the absence of ions and find a polarization profile as characterized in Figs. 4 and 5 of the main text. Here, we provide a more detailed discussion of the electrostatics of the system.

In order to develop a feel for the strength of the applied electric fields, we plot in Fig.~\ref{fig:polarization-SI} the $y$-component $P_y$ of the height-averaged polarization ${\bf P}=(P_x,P_y,0)$ along the $y$-axis, for several different electric field strengths. Clearly, under the twelve-fold increase of the electric field (from $0.25$ \ktea~ to $3$ \ktea), the polarization at the tip of the channel, where the field is strongest, increases by only a factor of roughly two. Thus, at these high electric fields water behaves as a gradually saturating non-linear dielectric, as it becomes increasingly difficult to further polarize water that is already strongly polarized. This differs from the linear low-field dielectric regime, where the polarization is proportional to the electric field \cite{Griffiths2017}.

\begin{figure}
    \centering
    \includegraphics[width=0.5\textwidth]{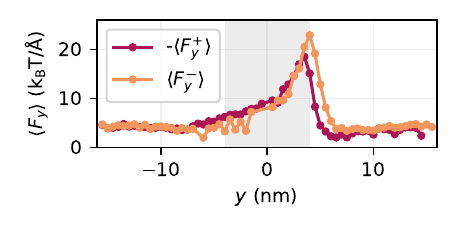}
    \caption{The forces on cations $-\langle F_y^+ \rangle$ and anions $\langle F_y^- \rangle$ located along the central axis, in the presence of another free ion pair, in a channel with tip-to-base width ratio $\xi=0.25$ and electric field strength of $E = -3$ k$_B$T/e$\rm \AA$, where it should be noted that $+\langle F_y^+ \rangle$ is negative for this electric field direction.}
    \label{fig:force_cb_extrasalt}
\end{figure}
\begin{figure}
    \centering
    \includegraphics[]{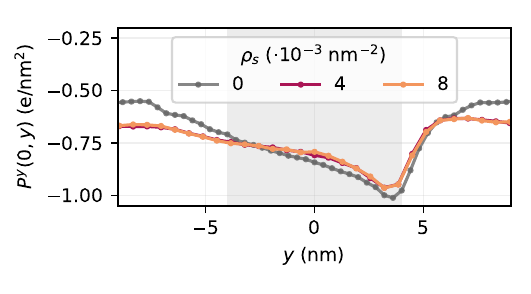}
    \caption{Polarization profile $P^y(0, y)$ along the central axis, in a channel with tip-to-base width ratio $\xi=0.25$ and electric field strength of $E = -3$ k$_B$T/e$\rm \AA$, averaged from constraint-biased simulations in the presence of an ion pair. The locations of the ions are changed along the central axis while keeping their distance fixed, as described above. The polarization profiles are averaged over all these configurations.}
    \label{fig:polarization_wsalt}
\end{figure}
As mentioned in the main text, it is not solely the $y$-component of ${\bf P}$ that leads to the formation of a divergence of ${\bf P}$, but for $E=+3$\ktea ~actually mostly the $x$-component perpendicular to the external field direction. Due to symmetry of the channel the $x$-component of the polarization vanishes on the central axis, $P_x(0,y)=0$, however this component changes sign on the central axis and hence $\partial_xP_x(0,y)\neq 0$ such that it contributes to $\nabla\cdot{\bf P}$ and hence to the bound charge.  In Fig.~\ref{fig:divergence-polarization-SI} we plot the derivatives $\partial_x P_x$ and $\partial_y P_y$ as well as their sum representing $\nabla\cdot{\bf P}$, similar to (the negative of) Fig.~4(e) in the main text, but now for (a) a strong, (b) an intermediate, and (c) a (relatively) weak electric field strength. We note that Fig.~\ref{fig:divergence-polarization-SI}(a) for $E=+3$~\ktea ~ presents the same data as in the main text,  showing opposite signs for $\partial_xP_x$ and $\partial_yP_y$ with the former term larger than the latter. For the weakest field $E=0.25$\ktea ~ of Fig.~\ref{fig:divergence-polarization-SI}(c), the two individual terms $\partial_x P_x$ and $\partial_y P_y$ also have opposite signs, however, they essentially add up to zero such that $\nabla\cdot{\bf P}$ and the bound charge become vanishingly small. The field strength of $E=1$\ktea ~ of Fig.~\ref{fig:divergence-polarization-SI}(b) shows an intermediate nonzero divergence of the polarization that is most prominent relatively close to the tip of the channel. Interestingly, we observe in Fig.~\ref{fig:divergence-polarization-SI} that $\partial_x P_x$ grows with $E$ (although sub-linearly) whereas $\partial_y P_y$ is essentially constant with $E$ and hence seemingly saturated, at least at the tip where it reaches the same minimum of $\partial_yP_y\simeq -0.2\text{e nm}^{-3}$ for all three field strengths. 

Finally, we stress that it is  the divergence of the polarization that causes the electric field-dependent ion selectivity, rather than just gradients of its components. Notably, even though there are substantial gradients in $P_x$ and $P_y$ throughout the channel at low electric field strengths, we do not observe significant ion selectivity under these conditions, as can also be seen in Fig. 2 of the main text, where $t_\pm\simeq 0.5$ for $|E|<0.5$\ktea. 

Many of the aforementioned analyses of  water polarization and the ion forces  were done in the low-salt limit, primarily due to the substantially lower computational cost in this dilute regime. However, salt can significantly impact the electrostatic behavior of this system, and it is not immediately clear whether our analysis remains valid at higher (finite) salt concentrations. An increase in salt concentration could lead to screening effects, potentially  altering the forces experienced by the ions. To compare the forces at different salt concentrations, we performed constraint-biased simulations where we calculated the average forces on a single ion pair, similar to our original system, but now also in the presence of a freely moving second ion pair. To increase statistics we performed these simulations for a very long time of $10$~ns, instead of the original $2$~ns, and disregard the force on an ion if one of the free ions is within a proximity of $1$~nm. Data points in which there are less than 2 valid time  steps (out of 1000) were discarded altogether. The result is shown in Fig. \ref{fig:force_cb_extrasalt}. Although the resulting data is still much noisier than in our original study, the results are almost identical to the ones presented in Fig. 5c of the main text. However, it is important to note that the original simulations were conducted at the opposite electric field.  These findings demonstrate that our results are not only valid in the infinitely dilute case, but extend to an ion concentration of at least $8\cdot 10^{-3}$~nm$^{-2}$, or 20 mM. Screening effects do not seem to significantly alter the forces acting on ions in this regime.

To verify that this also applies to water polarization, we  analyzed the water polarization in our constraint-biased calculations. Specifically, we examined  systems containing one fixed ion pair  (corresponding to  an areal density $\rho_s = 4\cdot 10^{-3}$~nm$^{-2}$), and systems with  one fixed ion pair along with an additional free ion pair (effectively at $\rho_s = 8\cdot 10^{-3}$~nm$^{-2}$). By averaging the water polarizations over all simulations with different fixed ions positions, we obtained the average water polarization in the presence of a modest salt concentration. This is shown in Fig. \ref{fig:polarization_wsalt}, along with the polarization profile in the absence of salt. The results show that the polarization profiles remain qualitatively similar and that the heterogeneity persists even at finite salt concentrations. There is a small discrepancy between the simulations with pure water and those with added salt, which resulted in slightly increased polarization in the reservoirs in the latter case. This is  likely due to the strong  local perturbations in the hydration layers of the fixed ions. However, we observe no significant  difference between the systems containing one and two ion pairs. Based on these findings, we conclude that, for the salt concentrations considered here, there is no clear evidence of significant screening effects in the water polarization profiles.

\section{Constraint-Biased Sampling}
\label{AppE}
In Fig.~5 of the main text we show the average forces acting on \xy ions at various positions along the channel. These forces were obtained using constraint-biased simulations,  in which a single cation and a single anion were included in the water system to ensure electro-neutrality. To minimize  their direct interaction, they were positioned far apart from each other. We fix one cation and one anion at a half-system distance of $16 \text{\,nm}$ from one another along the central axis of the channel. While water molecules are free to move, the ions are constrained to their respective positions.  After a short equilibration period of $20~\text{ps}$, we calculate the average forces acting on both ions due to the water and the external electric fields over a total simulation time of $2 \text{\,ns}$. This procedure is  repeated at various  positions along the entire central axis of our channel in steps of $5$ \angstrom, while keeping the distance between the ions fixed. This process  yields the average forces acting on both cations and anions throughout the channel.

\bibliography{library}

\begin{thebibliography}{46}%
\makeatletter
\providecommand \@ifxundefined [1]{%
 \@ifx{#1\undefined}
}%
\providecommand \@ifnum [1]{%
 \ifnum #1\expandafter \@firstoftwo
 \else \expandafter \@secondoftwo
 \fi
}%
\providecommand \@ifx [1]{%
 \ifx #1\expandafter \@firstoftwo
 \else \expandafter \@secondoftwo
 \fi
}%
\providecommand \natexlab [1]{#1}%
\providecommand \enquote  [1]{``#1''}%
\providecommand \bibnamefont  [1]{#1}%
\providecommand \bibfnamefont [1]{#1}%
\providecommand \citenamefont [1]{#1}%
\providecommand \href@noop [0]{\@secondoftwo}%
\providecommand \href [0]{\begingroup \@sanitize@url \@href}%
\providecommand \@href[1]{\@@startlink{#1}\@@href}%
\providecommand \@@href[1]{\endgroup#1\@@endlink}%
\providecommand \@sanitize@url [0]{\catcode `\\12\catcode `\$12\catcode
  `\&12\catcode `\#12\catcode `\^12\catcode `\_12\catcode `\%12\relax}%
\providecommand \@@startlink[1]{}%
\providecommand \@@endlink[0]{}%
\providecommand \url  [0]{\begingroup\@sanitize@url \@url }%
\providecommand \@url [1]{\endgroup\@href {#1}{\urlprefix }}%
\providecommand \urlprefix  [0]{URL }%
\providecommand \Eprint [0]{\href }%
\providecommand \doibase [0]{http://dx.doi.org/}%
\providecommand \selectlanguage [0]{\@gobble}%
\providecommand \bibinfo  [0]{\@secondoftwo}%
\providecommand \bibfield  [0]{\@secondoftwo}%
\providecommand \translation [1]{[#1]}%
\providecommand \BibitemOpen [0]{}%
\providecommand \bibitemStop [0]{}%
\providecommand \bibitemNoStop [0]{.\EOS\space}%
\providecommand \EOS [0]{\spacefactor3000\relax}%
\providecommand \BibitemShut  [1]{\csname bibitem#1\endcsname}%
\let\auto@bib@innerbib\@empty
\bibitem [{\citenamefont {Kavokine}\ \emph {et~al.}(2021)\citenamefont
  {Kavokine}, \citenamefont {Netz},\ and\ \citenamefont
  {Bocquet}}]{Kavokine2021}%
  \BibitemOpen
  \bibfield  {author} {\bibinfo {author} {\bibfnamefont {N.}~\bibnamefont
  {Kavokine}}, \bibinfo {author} {\bibfnamefont {R.~R.}\ \bibnamefont {Netz}},
  \ and\ \bibinfo {author} {\bibfnamefont {L.}~\bibnamefont {Bocquet}},\ }\href
  {\doibase 10.1146/ANNUREV-FLUID-071320-095958} {\bibfield  {journal}
  {\bibinfo  {journal} {Annu. Rev. Fluid Mech.}\ }\textbf {\bibinfo {volume}
  {53}},\ \bibinfo {pages} {377} (\bibinfo {year} {2021})}\BibitemShut
  {NoStop}%
\bibitem [{\citenamefont {Bocquet}(2020)}]{Bocquet2020}%
  \BibitemOpen
  \bibfield  {author} {\bibinfo {author} {\bibfnamefont {L.}~\bibnamefont
  {Bocquet}},\ }\href {\doibase 10.1038/S41563-020-0625-8} {\bibfield
  {journal} {\bibinfo  {journal} {Nat. Mater.}\ }\textbf {\bibinfo {volume}
  {19}},\ \bibinfo {pages} {254} (\bibinfo {year} {2020})}\BibitemShut
  {NoStop}%
\bibitem [{\citenamefont {Siwy}\ and\ \citenamefont
  {Fuli\'{n}ski}(2002)}]{Siwy20021}%
  \BibitemOpen
  \bibfield  {author} {\bibinfo {author} {\bibfnamefont {Z.}~\bibnamefont
  {Siwy}}\ and\ \bibinfo {author} {\bibfnamefont {A.}~\bibnamefont
  {Fuli\'{n}ski}},\ }\href {\doibase 10.1103/PhysRevLett.89.198103} {\bibfield
  {journal} {\bibinfo  {journal} {Phys. Rev. Lett.}\ }\textbf {\bibinfo
  {volume} {89}},\ \bibinfo {pages} {198103} (\bibinfo {year}
  {2002})}\BibitemShut {NoStop}%
\bibitem [{\citenamefont {Celebi}\ \emph {et~al.}(2014)\citenamefont {Celebi},
  \citenamefont {Buchheim}, \citenamefont {Wyss}, \citenamefont {Droudian},
  \citenamefont {Gasser}, \citenamefont {Shorubalko}, \citenamefont {Kye},
  \citenamefont {Lee},\ and\ \citenamefont {Park}}]{Celebi2014}%
  \BibitemOpen
  \bibfield  {author} {\bibinfo {author} {\bibfnamefont {K.}~\bibnamefont
  {Celebi}}, \bibinfo {author} {\bibfnamefont {J.}~\bibnamefont {Buchheim}},
  \bibinfo {author} {\bibfnamefont {R.~M.}\ \bibnamefont {Wyss}}, \bibinfo
  {author} {\bibfnamefont {A.}~\bibnamefont {Droudian}}, \bibinfo {author}
  {\bibfnamefont {P.}~\bibnamefont {Gasser}}, \bibinfo {author} {\bibfnamefont
  {I.}~\bibnamefont {Shorubalko}}, \bibinfo {author} {\bibfnamefont {J.~I.}\
  \bibnamefont {Kye}}, \bibinfo {author} {\bibfnamefont {C.}~\bibnamefont
  {Lee}}, \ and\ \bibinfo {author} {\bibfnamefont {H.~G.}\ \bibnamefont
  {Park}},\ }\href {\doibase 10.1126/SCIENCE.1249097} {\bibfield  {journal}
  {\bibinfo  {journal} {Science}\ }\textbf {\bibinfo {volume} {344}},\ \bibinfo
  {pages} {289} (\bibinfo {year} {2014})}\BibitemShut {NoStop}%
\bibitem [{\citenamefont {Garaj}\ \emph {et~al.}(2010)\citenamefont {Garaj},
  \citenamefont {Hubbard}, \citenamefont {Reina}, \citenamefont {Kong},
  \citenamefont {Branton},\ and\ \citenamefont {Golovchenko}}]{Garaj2010}%
  \BibitemOpen
  \bibfield  {author} {\bibinfo {author} {\bibfnamefont {S.}~\bibnamefont
  {Garaj}}, \bibinfo {author} {\bibfnamefont {W.}~\bibnamefont {Hubbard}},
  \bibinfo {author} {\bibfnamefont {A.}~\bibnamefont {Reina}}, \bibinfo
  {author} {\bibfnamefont {J.}~\bibnamefont {Kong}}, \bibinfo {author}
  {\bibfnamefont {D.}~\bibnamefont {Branton}}, \ and\ \bibinfo {author}
  {\bibfnamefont {J.~A.}\ \bibnamefont {Golovchenko}},\ }\href {\doibase
  10.1038/nature09379} {\bibfield  {journal} {\bibinfo  {journal} {Nature}\
  }\textbf {\bibinfo {volume} {467}},\ \bibinfo {pages} {190} (\bibinfo {year}
  {2010})}\BibitemShut {NoStop}%
\bibitem [{\citenamefont {Hummer}\ \emph {et~al.}(2001)\citenamefont {Hummer},
  \citenamefont {Rasaiah},\ and\ \citenamefont {Noworyta}}]{Hummer2001}%
  \BibitemOpen
  \bibfield  {author} {\bibinfo {author} {\bibfnamefont {G.}~\bibnamefont
  {Hummer}}, \bibinfo {author} {\bibfnamefont {J.~C.}\ \bibnamefont {Rasaiah}},
  \ and\ \bibinfo {author} {\bibfnamefont {J.~P.}\ \bibnamefont {Noworyta}},\
  }\href {\doibase 10.1038/35102535} {\bibfield  {journal} {\bibinfo  {journal}
  {Nature}\ }\textbf {\bibinfo {volume} {414}},\ \bibinfo {pages} {188}
  (\bibinfo {year} {2001})}\BibitemShut {NoStop}%
\bibitem [{\citenamefont {Esfandiar}\ \emph {et~al.}(2017)\citenamefont
  {Esfandiar}, \citenamefont {Radha}, \citenamefont {Wang}, \citenamefont
  {Yang}, \citenamefont {Hu}, \citenamefont {Garaj}, \citenamefont {Nair},
  \citenamefont {Geim},\ and\ \citenamefont {Gopinadhan}}]{Esfandiar2017}%
  \BibitemOpen
  \bibfield  {author} {\bibinfo {author} {\bibfnamefont {A.}~\bibnamefont
  {Esfandiar}}, \bibinfo {author} {\bibfnamefont {B.}~\bibnamefont {Radha}},
  \bibinfo {author} {\bibfnamefont {F.~C.}\ \bibnamefont {Wang}}, \bibinfo
  {author} {\bibfnamefont {Q.}~\bibnamefont {Yang}}, \bibinfo {author}
  {\bibfnamefont {S.}~\bibnamefont {Hu}}, \bibinfo {author} {\bibfnamefont
  {S.}~\bibnamefont {Garaj}}, \bibinfo {author} {\bibfnamefont {R.~R.}\
  \bibnamefont {Nair}}, \bibinfo {author} {\bibfnamefont {A.~K.}\ \bibnamefont
  {Geim}}, \ and\ \bibinfo {author} {\bibfnamefont {K.}~\bibnamefont
  {Gopinadhan}},\ }\href {\doibase 10.1126/SCIENCE.AAN5275} {\bibfield
  {journal} {\bibinfo  {journal} {Science}\ }\textbf {\bibinfo {volume}
  {358}},\ \bibinfo {pages} {511} (\bibinfo {year} {2017})}\BibitemShut
  {NoStop}%
\bibitem [{\citenamefont {Laucirica}\ \emph {et~al.}(2020)\citenamefont
  {Laucirica}, \citenamefont {Albesa}, \citenamefont {Toimil-Molares},
  \citenamefont {Trautmann}, \citenamefont {Marmisol\'{e}},\ and\ \citenamefont
  {Azzaroni}}]{Laucirica2020}%
  \BibitemOpen
  \bibfield  {author} {\bibinfo {author} {\bibfnamefont {G.}~\bibnamefont
  {Laucirica}}, \bibinfo {author} {\bibfnamefont {A.~G.}\ \bibnamefont
  {Albesa}}, \bibinfo {author} {\bibfnamefont {M.~E.}\ \bibnamefont
  {Toimil-Molares}}, \bibinfo {author} {\bibfnamefont {C.}~\bibnamefont
  {Trautmann}}, \bibinfo {author} {\bibfnamefont {W.~A.}\ \bibnamefont
  {Marmisol\'{e}}}, \ and\ \bibinfo {author} {\bibfnamefont {O.}~\bibnamefont
  {Azzaroni}},\ }\href {\doibase 10.1016/J.NANOEN.2020.104612} {\bibfield
  {journal} {\bibinfo  {journal} {Nano Energy}\ }\textbf {\bibinfo {volume}
  {71}},\ \bibinfo {pages} {104612} (\bibinfo {year} {2020})}\BibitemShut
  {NoStop}%
\bibitem [{\citenamefont {Siwy}\ \emph {et~al.}(2004)\citenamefont {Siwy},
  \citenamefont {Heins}, \citenamefont {Harrell}, \citenamefont {Kohli},\ and\
  \citenamefont {Martin}}]{Siwy2004}%
  \BibitemOpen
  \bibfield  {author} {\bibinfo {author} {\bibfnamefont {Z.}~\bibnamefont
  {Siwy}}, \bibinfo {author} {\bibfnamefont {E.}~\bibnamefont {Heins}},
  \bibinfo {author} {\bibfnamefont {C.~C.}\ \bibnamefont {Harrell}}, \bibinfo
  {author} {\bibfnamefont {P.}~\bibnamefont {Kohli}}, \ and\ \bibinfo {author}
  {\bibfnamefont {C.~R.}\ \bibnamefont {Martin}},\ }\href {\doibase
  10.1021/JA047675C} {\bibfield  {journal} {\bibinfo  {journal} {J. Am. Chem.
  Soc.}\ }\textbf {\bibinfo {volume} {126}},\ \bibinfo {pages} {10850}
  (\bibinfo {year} {2004})}\BibitemShut {NoStop}%
\bibitem [{\citenamefont {Stein}\ \emph {et~al.}(2004)\citenamefont {Stein},
  \citenamefont {Kruithof},\ and\ \citenamefont {Dekker}}]{Stein2004}%
  \BibitemOpen
  \bibfield  {author} {\bibinfo {author} {\bibfnamefont {D.}~\bibnamefont
  {Stein}}, \bibinfo {author} {\bibfnamefont {M.}~\bibnamefont {Kruithof}}, \
  and\ \bibinfo {author} {\bibfnamefont {C.}~\bibnamefont {Dekker}},\ }\href
  {\doibase 10.1103/PHYSREVLETT.93.035901} {\bibfield  {journal} {\bibinfo
  {journal} {Phys. Rev. Lett.}\ }\textbf {\bibinfo {volume} {93}},\ \bibinfo
  {pages} {035901} (\bibinfo {year} {2004})}\BibitemShut {NoStop}%
\bibitem [{\citenamefont {Wei}\ \emph {et~al.}(1997)\citenamefont {Wei},
  \citenamefont {Bard},\ and\ \citenamefont {Feldberg}}]{Wei1997}%
  \BibitemOpen
  \bibfield  {author} {\bibinfo {author} {\bibfnamefont {C.}~\bibnamefont
  {Wei}}, \bibinfo {author} {\bibfnamefont {A.~J.}\ \bibnamefont {Bard}}, \
  and\ \bibinfo {author} {\bibfnamefont {S.~W.}\ \bibnamefont {Feldberg}},\
  }\href {\doibase 10.1021/AC970551G} {\bibfield  {journal} {\bibinfo
  {journal} {Anal. Chem.}\ }\textbf {\bibinfo {volume} {69}},\ \bibinfo {pages}
  {4627} (\bibinfo {year} {1997})}\BibitemShut {NoStop}%
\bibitem [{\citenamefont {Siwy}\ \emph {et~al.}(2002)\citenamefont {Siwy},
  \citenamefont {Gu}, \citenamefont {Spohr}, \citenamefont {Baur},
  \citenamefont {Wolf-Reber}, \citenamefont {Spohr}, \citenamefont {Apel},\
  and\ \citenamefont {Korchev}}]{Siwy20022}%
  \BibitemOpen
  \bibfield  {author} {\bibinfo {author} {\bibfnamefont {Z.}~\bibnamefont
  {Siwy}}, \bibinfo {author} {\bibfnamefont {Y.}~\bibnamefont {Gu}}, \bibinfo
  {author} {\bibfnamefont {H.~A.}\ \bibnamefont {Spohr}}, \bibinfo {author}
  {\bibfnamefont {D.}~\bibnamefont {Baur}}, \bibinfo {author} {\bibfnamefont
  {A.}~\bibnamefont {Wolf-Reber}}, \bibinfo {author} {\bibfnamefont
  {R.}~\bibnamefont {Spohr}}, \bibinfo {author} {\bibfnamefont
  {P.}~\bibnamefont {Apel}}, \ and\ \bibinfo {author} {\bibfnamefont {Y.~E.}\
  \bibnamefont {Korchev}},\ }\href {\doibase 10.1209/EPL/I2002-00271-3}
  {\bibfield  {journal} {\bibinfo  {journal} {Europhys. Lett.}\ }\textbf
  {\bibinfo {volume} {60}},\ \bibinfo {pages} {349} (\bibinfo {year}
  {2002})}\BibitemShut {NoStop}%
\bibitem [{\citenamefont {Siwy}(2006)}]{Siwy2006}%
  \BibitemOpen
  \bibfield  {author} {\bibinfo {author} {\bibfnamefont {Z.~S.}\ \bibnamefont
  {Siwy}},\ }\href {\doibase 10.1002/ADFM.200500471} {\bibfield  {journal}
  {\bibinfo  {journal} {Adv. Funct. Mater.}\ }\textbf {\bibinfo {volume}
  {16}},\ \bibinfo {pages} {735} (\bibinfo {year} {2006})}\BibitemShut
  {NoStop}%
\bibitem [{\citenamefont {Karnik}\ \emph {et~al.}(2007)\citenamefont {Karnik},
  \citenamefont {Duan}, \citenamefont {Castelino}, \citenamefont {Daiguji},\
  and\ \citenamefont {Majumdar}}]{Karnik2007}%
  \BibitemOpen
  \bibfield  {author} {\bibinfo {author} {\bibfnamefont {R.}~\bibnamefont
  {Karnik}}, \bibinfo {author} {\bibfnamefont {C.}~\bibnamefont {Duan}},
  \bibinfo {author} {\bibfnamefont {K.}~\bibnamefont {Castelino}}, \bibinfo
  {author} {\bibfnamefont {H.}~\bibnamefont {Daiguji}}, \ and\ \bibinfo
  {author} {\bibfnamefont {A.}~\bibnamefont {Majumdar}},\ }\href {\doibase
  10.1021/NL062806O} {\bibfield  {journal} {\bibinfo  {journal} {Nano Lett.}\
  }\textbf {\bibinfo {volume} {7}},\ \bibinfo {pages} {547} (\bibinfo {year}
  {2007})}\BibitemShut {NoStop}%
\bibitem [{\citenamefont {Jubin}\ \emph {et~al.}(2018)\citenamefont {Jubin},
  \citenamefont {Poggioli}, \citenamefont {Siria},\ and\ \citenamefont
  {Bocquet}}]{Jubin2018}%
  \BibitemOpen
  \bibfield  {author} {\bibinfo {author} {\bibfnamefont {L.}~\bibnamefont
  {Jubin}}, \bibinfo {author} {\bibfnamefont {A.}~\bibnamefont {Poggioli}},
  \bibinfo {author} {\bibfnamefont {A.}~\bibnamefont {Siria}}, \ and\ \bibinfo
  {author} {\bibfnamefont {L.}~\bibnamefont {Bocquet}},\ }\href {\doibase
  10.1073/pnas.1721987115} {\bibfield  {journal} {\bibinfo  {journal} {Proc.
  Natl. Acad. Sci. U. S. A.}\ }\textbf {\bibinfo {volume} {115}},\ \bibinfo
  {pages} {4063} (\bibinfo {year} {2018})}\BibitemShut {NoStop}%
\bibitem [{\citenamefont {Boon}\ \emph {et~al.}(2022)\citenamefont {Boon},
  \citenamefont {Veenstra}, \citenamefont {Dijkstra},\ and\ \citenamefont {van
  Roij}}]{Boon2022}%
  \BibitemOpen
  \bibfield  {author} {\bibinfo {author} {\bibfnamefont {W.~Q.}\ \bibnamefont
  {Boon}}, \bibinfo {author} {\bibfnamefont {T.~E.}\ \bibnamefont {Veenstra}},
  \bibinfo {author} {\bibfnamefont {M.}~\bibnamefont {Dijkstra}}, \ and\
  \bibinfo {author} {\bibfnamefont {R.}~\bibnamefont {van Roij}},\ }\href
  {\doibase 10.1063/5.0113035} {\bibfield  {journal} {\bibinfo  {journal}
  {Phys. Fluids}\ }\textbf {\bibinfo {volume} {34}},\ \bibinfo {pages} {101701}
  (\bibinfo {year} {2022})}\BibitemShut {NoStop}%
\bibitem [{\citenamefont {Novoselov}\ \emph {et~al.}(2004)\citenamefont
  {Novoselov}, \citenamefont {Geim}, \citenamefont {Morozov}, \citenamefont
  {Jiang}, \citenamefont {Zhang}, \citenamefont {Dubonos}, \citenamefont
  {Grigorieva},\ and\ \citenamefont {Firsov}}]{Novoselov2004}%
  \BibitemOpen
  \bibfield  {author} {\bibinfo {author} {\bibfnamefont {K.~S.}\ \bibnamefont
  {Novoselov}}, \bibinfo {author} {\bibfnamefont {A.~K.}\ \bibnamefont {Geim}},
  \bibinfo {author} {\bibfnamefont {S.~V.}\ \bibnamefont {Morozov}}, \bibinfo
  {author} {\bibfnamefont {D.}~\bibnamefont {Jiang}}, \bibinfo {author}
  {\bibfnamefont {Y.}~\bibnamefont {Zhang}}, \bibinfo {author} {\bibfnamefont
  {S.~V.}\ \bibnamefont {Dubonos}}, \bibinfo {author} {\bibfnamefont {I.~V.}\
  \bibnamefont {Grigorieva}}, \ and\ \bibinfo {author} {\bibfnamefont {A.~A.}\
  \bibnamefont {Firsov}},\ }\href {\doibase 10.1126/SCIENCE.1102896} {\bibfield
   {journal} {\bibinfo  {journal} {Science}\ }\textbf {\bibinfo {volume}
  {306}},\ \bibinfo {pages} {666} (\bibinfo {year} {2004})}\BibitemShut
  {NoStop}%
\bibitem [{\citenamefont {Keerthi}\ \emph {et~al.}(2018)\citenamefont
  {Keerthi}, \citenamefont {Geim}, \citenamefont {Janardanan}, \citenamefont
  {Rooney}, \citenamefont {Esfandiar}, \citenamefont {Hu}, \citenamefont {Dar},
  \citenamefont {Grigorieva}, \citenamefont {Haigh}, \citenamefont {Wang},\
  and\ \citenamefont {Radha}}]{Keerthi2018}%
  \BibitemOpen
  \bibfield  {author} {\bibinfo {author} {\bibfnamefont {A.}~\bibnamefont
  {Keerthi}}, \bibinfo {author} {\bibfnamefont {A.~K.}\ \bibnamefont {Geim}},
  \bibinfo {author} {\bibfnamefont {A.}~\bibnamefont {Janardanan}}, \bibinfo
  {author} {\bibfnamefont {A.~P.}\ \bibnamefont {Rooney}}, \bibinfo {author}
  {\bibfnamefont {A.}~\bibnamefont {Esfandiar}}, \bibinfo {author}
  {\bibfnamefont {S.}~\bibnamefont {Hu}}, \bibinfo {author} {\bibfnamefont
  {S.~A.}\ \bibnamefont {Dar}}, \bibinfo {author} {\bibfnamefont {I.~V.}\
  \bibnamefont {Grigorieva}}, \bibinfo {author} {\bibfnamefont {S.~J.}\
  \bibnamefont {Haigh}}, \bibinfo {author} {\bibfnamefont {F.~C.}\ \bibnamefont
  {Wang}}, \ and\ \bibinfo {author} {\bibfnamefont {B.}~\bibnamefont {Radha}},\
  }\href {\doibase 10.1038/S41586-018-0203-2} {\bibfield  {journal} {\bibinfo
  {journal} {Nature}\ }\textbf {\bibinfo {volume} {558}},\ \bibinfo {pages}
  {420} (\bibinfo {year} {2018})}\BibitemShut {NoStop}%
\bibitem [{\citenamefont {Geim}(2021)}]{Geim2021}%
  \BibitemOpen
  \bibfield  {author} {\bibinfo {author} {\bibfnamefont {A.~K.}\ \bibnamefont
  {Geim}},\ }\href {\doibase 10.1021/ACS.NANOLETT.1C02591} {\bibfield
  {journal} {\bibinfo  {journal} {Nano Lett.}\ }\textbf {\bibinfo {volume}
  {21}},\ \bibinfo {pages} {6356} (\bibinfo {year} {2021})}\BibitemShut
  {NoStop}%
\bibitem [{\citenamefont {Fumagalli}\ \emph {et~al.}(2018)\citenamefont
  {Fumagalli}, \citenamefont {Esfandiar}, \citenamefont {Fabregas},
  \citenamefont {Hu}, \citenamefont {Ares}, \citenamefont {Janardanan},
  \citenamefont {Yang}, \citenamefont {Radha}, \citenamefont {Taniguchi},
  \citenamefont {Watanabe}, \citenamefont {Gomila}, \citenamefont {Novoselov},\
  and\ \citenamefont {Geim}}]{Fumagalli2018}%
  \BibitemOpen
  \bibfield  {author} {\bibinfo {author} {\bibfnamefont {L.}~\bibnamefont
  {Fumagalli}}, \bibinfo {author} {\bibfnamefont {A.}~\bibnamefont
  {Esfandiar}}, \bibinfo {author} {\bibfnamefont {R.}~\bibnamefont {Fabregas}},
  \bibinfo {author} {\bibfnamefont {S.}~\bibnamefont {Hu}}, \bibinfo {author}
  {\bibfnamefont {P.}~\bibnamefont {Ares}}, \bibinfo {author} {\bibfnamefont
  {A.}~\bibnamefont {Janardanan}}, \bibinfo {author} {\bibfnamefont
  {Q.}~\bibnamefont {Yang}}, \bibinfo {author} {\bibfnamefont {B.}~\bibnamefont
  {Radha}}, \bibinfo {author} {\bibfnamefont {T.}~\bibnamefont {Taniguchi}},
  \bibinfo {author} {\bibfnamefont {K.}~\bibnamefont {Watanabe}}, \bibinfo
  {author} {\bibfnamefont {G.}~\bibnamefont {Gomila}}, \bibinfo {author}
  {\bibfnamefont {K.~S.}\ \bibnamefont {Novoselov}}, \ and\ \bibinfo {author}
  {\bibfnamefont {A.~K.}\ \bibnamefont {Geim}},\ }\href {\doibase
  10.1126/SCIENCE.AAT4191} {\bibfield  {journal} {\bibinfo  {journal}
  {Science}\ }\textbf {\bibinfo {volume} {360}},\ \bibinfo {pages} {1339}
  (\bibinfo {year} {2018})}\BibitemShut {NoStop}%
\bibitem [{\citenamefont {Robin}\ \emph {et~al.}(2021)\citenamefont {Robin},
  \citenamefont {Kavokine},\ and\ \citenamefont {Bocquet}}]{Robin2021}%
  
  \BibitemOpen
  \bibfield  {author} 
  {\bibinfo {author} {\bibfnamefont {P.}~\bibnamefont
  {Robin}}, 
  \bibinfo {author} {\bibfnamefont {N.}~\bibnamefont {Kavokine}}, \
  and\ \bibinfo {author} {\bibfnamefont {L.}~\bibnamefont {Bocquet}},\ }
  
  \href {\doibase 10.1126/SCIENCE.ABF7923} 
  {\bibfield  {journal} {\bibinfo  {journal}
  {Science}\ }\textbf {\bibinfo {volume} {373}},\ \bibinfo {pages} {687}
  (\bibinfo {year} {2021})}
  \BibitemShut {NoStop}%
\bibitem [{\citenamefont {Zhao}\ \emph {et~al.}(2021)\citenamefont {Zhao},
  \citenamefont {Sun}, \citenamefont {Zhu}, \citenamefont {Jiang},
  \citenamefont {Zhao}, \citenamefont {Lin}, \citenamefont {Xu}, \citenamefont
  {Duan}, \citenamefont {Francisco},\ and\ \citenamefont {Zeng}}]{Zhao2021}%
  \BibitemOpen
  \bibfield  {author} {\bibinfo {author} {\bibfnamefont {W.}~\bibnamefont
  {Zhao}}, \bibinfo {author} {\bibfnamefont {Y.}~\bibnamefont {Sun}}, \bibinfo
  {author} {\bibfnamefont {W.}~\bibnamefont {Zhu}}, \bibinfo {author}
  {\bibfnamefont {J.}~\bibnamefont {Jiang}}, \bibinfo {author} {\bibfnamefont
  {X.}~\bibnamefont {Zhao}}, \bibinfo {author} {\bibfnamefont {D.}~\bibnamefont
  {Lin}}, \bibinfo {author} {\bibfnamefont {W.}~\bibnamefont {Xu}}, \bibinfo
  {author} {\bibfnamefont {X.}~\bibnamefont {Duan}}, \bibinfo {author}
  {\bibfnamefont {J.~S.}\ \bibnamefont {Francisco}}, \ and\ \bibinfo {author}
  {\bibfnamefont {X.~C.}\ \bibnamefont {Zeng}},\ }\href {\doibase
  10.1038/s41467-021-25938-0} {\bibfield  {journal} {\bibinfo  {journal} {Nat.
  Commun.}\ }\textbf {\bibinfo {volume} {12}},\ \bibinfo {pages} {1} (\bibinfo
  {year} {2021})}\BibitemShut {NoStop}%
\bibitem [{\citenamefont {Robin}\ \emph {et~al.}(2023)\citenamefont {Robin},
  \citenamefont {Emmerich}, \citenamefont {Ismail}, \citenamefont {Nigu\`{e}s},
  \citenamefont {You}, \citenamefont {Nam}, \citenamefont {Keerthi},
  \citenamefont {Siria}, \citenamefont {Geim}, \citenamefont {Radha},\ and\
  \citenamefont {Bocquet}}]{Robin2023}%
  \BibitemOpen
  \bibfield  {author} {\bibinfo {author} {\bibfnamefont {P.}~\bibnamefont
  {Robin}}, \bibinfo {author} {\bibfnamefont {T.}~\bibnamefont {Emmerich}},
  \bibinfo {author} {\bibfnamefont {A.}~\bibnamefont {Ismail}}, \bibinfo
  {author} {\bibfnamefont {A.}~\bibnamefont {Nigu\`{e}s}}, \bibinfo {author}
  {\bibfnamefont {Y.}~\bibnamefont {You}}, \bibinfo {author} {\bibfnamefont
  {G.-H.}\ \bibnamefont {Nam}}, \bibinfo {author} {\bibfnamefont
  {A.}~\bibnamefont {Keerthi}}, \bibinfo {author} {\bibfnamefont
  {A.}~\bibnamefont {Siria}}, \bibinfo {author} {\bibfnamefont {A.~K.}\
  \bibnamefont {Geim}}, \bibinfo {author} {\bibfnamefont {B.}~\bibnamefont
  {Radha}}, \ and\ \bibinfo {author} {\bibfnamefont {L.}~\bibnamefont
  {Bocquet}},\ }\href {\doibase 10.1126/science.adc9931} {\bibfield  {journal}
  {\bibinfo  {journal} {Science}\ }\textbf {\bibinfo {volume} {379}},\ \bibinfo
  {pages} {161} (\bibinfo {year} {2023})}\BibitemShut {NoStop}%
\bibitem [{\citenamefont {Fornasiero}\ \emph {et~al.}(2008)\citenamefont
  {Fornasiero}, \citenamefont {Hyung}, \citenamefont {Holt}, \citenamefont
  {Stadermann}, \citenamefont {Grigoropoulos}, \citenamefont {Noy},\ and\
  \citenamefont {Bakajin}}]{Fornasiero2008}%
  \BibitemOpen
  \bibfield  {author} {\bibinfo {author} {\bibfnamefont {F.}~\bibnamefont
  {Fornasiero}}, \bibinfo {author} {\bibfnamefont {G.~P.}\ \bibnamefont
  {Hyung}}, \bibinfo {author} {\bibfnamefont {J.~K.}\ \bibnamefont {Holt}},
  \bibinfo {author} {\bibfnamefont {M.}~\bibnamefont {Stadermann}}, \bibinfo
  {author} {\bibfnamefont {C.~P.}\ \bibnamefont {Grigoropoulos}}, \bibinfo
  {author} {\bibfnamefont {A.}~\bibnamefont {Noy}}, \ and\ \bibinfo {author}
  {\bibfnamefont {O.}~\bibnamefont {Bakajin}},\ }\href {\doibase
  10.1073/pnas.0710437105} {\bibfield  {journal} {\bibinfo  {journal} {Proc.
  Natl. Acad. Sci. USA.}\ }\textbf {\bibinfo {volume} {105}},\ \bibinfo {pages}
  {17250} (\bibinfo {year} {2008})}\BibitemShut {NoStop}%
\bibitem [{\citenamefont {Yu}\ \emph {et~al.}(2019)\citenamefont {Yu},
  \citenamefont {Fan}, \citenamefont {Xia}, \citenamefont {Zhu}, \citenamefont
  {Wu},\ and\ \citenamefont {Wang}}]{Yu2019}%
  \BibitemOpen
  \bibfield  {author} {\bibinfo {author} {\bibfnamefont {Y.}~\bibnamefont
  {Yu}}, \bibinfo {author} {\bibfnamefont {J.}~\bibnamefont {Fan}}, \bibinfo
  {author} {\bibfnamefont {J.}~\bibnamefont {Xia}}, \bibinfo {author}
  {\bibfnamefont {Y.}~\bibnamefont {Zhu}}, \bibinfo {author} {\bibfnamefont
  {H.}~\bibnamefont {Wu}}, \ and\ \bibinfo {author} {\bibfnamefont
  {F.}~\bibnamefont {Wang}},\ }\href {\doibase 10.1039/C9NR00317G} {\bibfield
  {journal} {\bibinfo  {journal} {Nanoscale}\ }\textbf {\bibinfo {volume}
  {11}},\ \bibinfo {pages} {8449} (\bibinfo {year} {2019})}\BibitemShut
  {NoStop}%
\bibitem [{\citenamefont {Sahu}\ and\ \citenamefont {Zwolak}(2017)}]{Sahu2017}%
  \BibitemOpen
  \bibfield  {author} {\bibinfo {author} {\bibfnamefont {S.}~\bibnamefont
  {Sahu}}\ and\ \bibinfo {author} {\bibfnamefont {M.}~\bibnamefont {Zwolak}},\
  }\href {\doibase 10.1039/C7NR03838K} {\bibfield  {journal} {\bibinfo
  {journal} {Nanoscale}\ }\textbf {\bibinfo {volume} {9}},\ \bibinfo {pages}
  {11424} (\bibinfo {year} {2017})}\BibitemShut {NoStop}%
\bibitem [{\citenamefont {Xue}\ \emph {et~al.}(2022)\citenamefont {Xue},
  \citenamefont {Qiu}, \citenamefont {Shen}, \citenamefont {Zhang},\ and\
  \citenamefont {Guo}}]{Xue2022}%
  \BibitemOpen
  \bibfield  {author} {\bibinfo {author} {\bibfnamefont {M.}~\bibnamefont
  {Xue}}, \bibinfo {author} {\bibfnamefont {H.}~\bibnamefont {Qiu}}, \bibinfo
  {author} {\bibfnamefont {C.}~\bibnamefont {Shen}}, \bibinfo {author}
  {\bibfnamefont {Z.}~\bibnamefont {Zhang}}, \ and\ \bibinfo {author}
  {\bibfnamefont {W.}~\bibnamefont {Guo}},\ }\href {\doibase
  10.1021/ACS.JPCLETT.2C00817} {\bibfield  {journal} {\bibinfo  {journal} {J.
  Phys. Chem. Lett.}\ }\textbf {\bibinfo {volume} {13}},\ \bibinfo {pages}
  {4815} (\bibinfo {year} {2022})}\BibitemShut {NoStop}%
\bibitem [{\citenamefont {Li}\ \emph {et~al.}(2021)\citenamefont {Li},
  \citenamefont {Zhao}, \citenamefont {Zhang}, \citenamefont {Ding},\ and\
  \citenamefont {Su}}]{Li2021}%
  \BibitemOpen
  \bibfield  {author} {\bibinfo {author} {\bibfnamefont {S.}~\bibnamefont
  {Li}}, \bibinfo {author} {\bibfnamefont {Y.}~\bibnamefont {Zhao}}, \bibinfo
  {author} {\bibfnamefont {X.}~\bibnamefont {Zhang}}, \bibinfo {author}
  {\bibfnamefont {C.}~\bibnamefont {Ding}}, \ and\ \bibinfo {author}
  {\bibfnamefont {J.}~\bibnamefont {Su}},\ }\href {\doibase
  10.1021/ACS.JPCB.1C05255} {\bibfield  {journal} {\bibinfo  {journal} {J.
  Phys. Chem. B}\ }\textbf {\bibinfo {volume} {125}},\ \bibinfo {pages} {11232}
  (\bibinfo {year} {2021})}\BibitemShut {NoStop}%
\bibitem [{\citenamefont {Li}\ \emph {et~al.}(2022)\citenamefont {Li},
  \citenamefont {Zhang}, \citenamefont {Liu},\ and\ \citenamefont
  {Su}}]{Li2022b}%
  \BibitemOpen
  \bibfield  {author} {\bibinfo {author} {\bibfnamefont {S.}~\bibnamefont
  {Li}}, \bibinfo {author} {\bibfnamefont {X.}~\bibnamefont {Zhang}}, \bibinfo
  {author} {\bibfnamefont {Y.}~\bibnamefont {Liu}}, \ and\ \bibinfo {author}
  {\bibfnamefont {J.}~\bibnamefont {Su}},\ }\href {\doibase 10.1039/D2CP00025C}
  {\bibfield  {journal} {\bibinfo  {journal} {Phys. Chem. Chem. Phys.}\
  }\textbf {\bibinfo {volume} {24}},\ \bibinfo {pages} {13245} (\bibinfo {year}
  {2022})}\BibitemShut {NoStop}%
\bibitem [{\citenamefont {Zhou}\ \emph {et~al.}(2023)\citenamefont {Zhou},
  \citenamefont {Guo}, \citenamefont {Zhang}, \citenamefont {Guo},\ and\
  \citenamefont {Qiu}}]{Zhou2023}%
  \BibitemOpen
  \bibfield  {author} {\bibinfo {author} {\bibfnamefont {W.}~\bibnamefont
  {Zhou}}, \bibinfo {author} {\bibfnamefont {Y.}~\bibnamefont {Guo}}, \bibinfo
  {author} {\bibfnamefont {Z.}~\bibnamefont {Zhang}}, \bibinfo {author}
  {\bibfnamefont {W.}~\bibnamefont {Guo}}, \ and\ \bibinfo {author}
  {\bibfnamefont {H.}~\bibnamefont {Qiu}},\ }\href {\doibase
  10.1103/PhysRevLett.130.084001} {\bibfield  {journal} {\bibinfo  {journal}
  {Phys. Rev. Lett.}\ }\textbf {\bibinfo {volume} {130}},\ \bibinfo {pages}
  {084001} (\bibinfo {year} {2023})}\BibitemShut {NoStop}%
\bibitem [{\citenamefont {Siria}\ \emph {et~al.}(2017)\citenamefont {Siria},
  \citenamefont {Bocquet},\ and\ \citenamefont {Bocquet}}]{Siria2017}%
  \BibitemOpen
  \bibfield  {author} {\bibinfo {author} {\bibfnamefont {A.}~\bibnamefont
  {Siria}}, \bibinfo {author} {\bibfnamefont {M.~L.}\ \bibnamefont {Bocquet}},
  \ and\ \bibinfo {author} {\bibfnamefont {L.}~\bibnamefont {Bocquet}},\ }\href
  {\doibase 10.1038/S41570-017-0091} {\bibfield  {journal} {\bibinfo  {journal}
  {Nat. Rev. Chem.}\ }\textbf {\bibinfo {volume} {1}},\ \bibinfo {pages} {1}
  (\bibinfo {year} {2017})}\BibitemShut {NoStop}%
\bibitem [{\citenamefont {Werber}\ \emph {et~al.}(2016)\citenamefont {Werber},
  \citenamefont {Osuji},\ and\ \citenamefont {Elimelech}}]{Werber2016}%
  \BibitemOpen
  \bibfield  {author} {\bibinfo {author} {\bibfnamefont {J.~R.}\ \bibnamefont
  {Werber}}, \bibinfo {author} {\bibfnamefont {C.~O.}\ \bibnamefont {Osuji}}, \
  and\ \bibinfo {author} {\bibfnamefont {M.}~\bibnamefont {Elimelech}},\ }\href
  {\doibase 10.1038/natrevmats.2016.18} {\bibfield  {journal} {\bibinfo
  {journal} {Nat. Rev. Mater.}\ }\textbf {\bibinfo {volume} {1}},\ \bibinfo
  {pages} {1} (\bibinfo {year} {2016})}\BibitemShut {NoStop}%
\bibitem [{\citenamefont {Kamsma}\ \emph {et~al.}(2023)\citenamefont {Kamsma},
  \citenamefont {Boon}, \citenamefont {ter Rele}, \citenamefont {Spitoni},\
  and\ \citenamefont {van Roij}}]{Kamsma2023}%
  \BibitemOpen
  \bibfield  {author} {\bibinfo {author} {\bibfnamefont {T.~M.}\ \bibnamefont
  {Kamsma}}, \bibinfo {author} {\bibfnamefont {W.~Q.}\ \bibnamefont {Boon}},
  \bibinfo {author} {\bibfnamefont {T.}~\bibnamefont {ter Rele}}, \bibinfo
  {author} {\bibfnamefont {C.}~\bibnamefont {Spitoni}}, \ and\ \bibinfo
  {author} {\bibfnamefont {R.}~\bibnamefont {van Roij}},\ }\href {\doibase
  10.1103/PhysRevLett.130.268401} {\bibfield  {journal} {\bibinfo  {journal}
  {Phys. Rev. Lett.}\ }\textbf {\bibinfo {volume} {130}},\ \bibinfo {pages}
  {268401} (\bibinfo {year} {2023})}\BibitemShut {NoStop}%
\bibitem [{\citenamefont {Yoshida}\ \emph {et~al.}(2018)\citenamefont
  {Yoshida}, \citenamefont {Kaiser}, \citenamefont {Rotenberg},\ and\
  \citenamefont {Bocquet}}]{Yoshida2018}%
  \BibitemOpen
  \bibfield  {author} {\bibinfo {author} {\bibfnamefont {H.}~\bibnamefont
  {Yoshida}}, \bibinfo {author} {\bibfnamefont {V.}~\bibnamefont {Kaiser}},
  \bibinfo {author} {\bibfnamefont {B.}~\bibnamefont {Rotenberg}}, \ and\
  \bibinfo {author} {\bibfnamefont {L.}~\bibnamefont {Bocquet}},\ }\href
  {\doibase 10.1038/s41467-018-03829-1} {\bibfield  {journal} {\bibinfo
  {journal} {Nat. Commun.}\ }\textbf {\bibinfo {volume} {9}},\ \bibinfo {pages}
  {1} (\bibinfo {year} {2018})}\BibitemShut {NoStop}%
\bibitem [{\citenamefont {Berendsen}\ \emph {et~al.}(2002)\citenamefont
  {Berendsen}, \citenamefont {Grigera},\ and\ \citenamefont
  {Straatsma}}]{Berendsen2002}%
  \BibitemOpen
  \bibfield  {author} {\bibinfo {author} {\bibfnamefont {H.~J.}\ \bibnamefont
  {Berendsen}}, \bibinfo {author} {\bibfnamefont {J.~R.}\ \bibnamefont
  {Grigera}}, \ and\ \bibinfo {author} {\bibfnamefont {T.~P.}\ \bibnamefont
  {Straatsma}},\ }\href {\doibase 10.1021/J100308A038} {\bibfield  {journal}
  {\bibinfo  {journal} {J. Phys. Chem.}\ }\textbf {\bibinfo {volume} {91}},\
  \bibinfo {pages} {6269} (\bibinfo {year} {2002})}\BibitemShut {NoStop}%
\bibitem [{\citenamefont {Atkins}\ \emph {et~al.}(2018)\citenamefont {Atkins},
  \citenamefont {De~Paula},\ and\ \citenamefont {Keeler}}]{Atkins2018}%
  \BibitemOpen
  \bibfield  {author} {\bibinfo {author} {\bibfnamefont {P.}~\bibnamefont
  {Atkins}}, \bibinfo {author} {\bibfnamefont {J.}~\bibnamefont {De~Paula}}, \
  and\ \bibinfo {author} {\bibfnamefont {J.}~\bibnamefont {Keeler}},\
  }\href@noop {} {\emph {\bibinfo {title} {Atkins' Physical Chemistry}}}\
  (\bibinfo  {publisher} {Oxford University Press},\ \bibinfo {year}
  {2018})\BibitemShut {NoStop}%
\bibitem [{\citenamefont {Delgado}\ \emph {et~al.}(2005)\citenamefont
  {Delgado}, \citenamefont {Gonz\'alez-Caballero}, \citenamefont {Hunter},
  \citenamefont {Koopal},\ and\ \citenamefont {Lyklema}}]{Delgado2005}%
  \BibitemOpen
  \bibfield  {author} {\bibinfo {author} {\bibfnamefont {A.~V.}\ \bibnamefont
  {Delgado}}, \bibinfo {author} {\bibfnamefont {F.}~\bibnamefont
  {Gonz\'alez-Caballero}}, \bibinfo {author} {\bibfnamefont {R.~J.}\
  \bibnamefont {Hunter}}, \bibinfo {author} {\bibfnamefont {L.~K.}\
  \bibnamefont {Koopal}}, \ and\ \bibinfo {author} {\bibfnamefont
  {J.}~\bibnamefont {Lyklema}},\ }\href {\doibase doi:10.1351/pac200577101753}
  {\bibfield  {journal} {\bibinfo  {journal} {Pure Appl. Chem.}\ }\textbf
  {\bibinfo {volume} {77}},\ \bibinfo {pages} {1753} (\bibinfo {year}
  {2005})}\BibitemShut {NoStop}%
\bibitem [{\citenamefont {Griffiths}(2017)}]{Griffiths2017}%
  \BibitemOpen
  \bibfield  {author} {\bibinfo {author} {\bibfnamefont {D.~J.}\ \bibnamefont
  {Griffiths}},\ }\href {\doibase 10.1017/9781108333511} {\emph {\bibinfo
  {title} {Introduction to Electrodynamics}}}\ (\bibinfo  {publisher}
  {Cambridge University Press},\ \bibinfo {year} {2017})\BibitemShut {NoStop}%
\bibitem [{\citenamefont {Maggs}\ \emph {et~al.}(2006)\citenamefont {Maggs},
  \citenamefont {Maggs},\ and\ \citenamefont
  {Everaers}}]{Maggs2006}%
  \BibitemOpen
  \bibfield  {author} {\bibinfo {author} {\bibfnamefont {A.C.}~\bibnamefont
  {Maggs}}  and\
  \bibinfo {author} {\bibfnamefont {R.}~\bibnamefont {Everaers}},\ }\href
  {\doibase https://doi.org/10.1103/PhysRevLett.96.230603} {\bibfield
  {journal} {\bibinfo  {journal} {Phys. Rev. Lett.}\ }\textbf {\bibinfo
  {volume} {96}},\ \bibinfo {pages} {230603} (\bibinfo {year}
  {2006})}\BibitemShut {NoStop}%
\bibitem [{\citenamefont {Paillusson}\ \emph {et~al.}(2010)\citenamefont {Paillusson},
  \citenamefont {Paillusson} and\ \citenamefont
  {Blossey}}]{Paillusson2010}%
  \BibitemOpen
  \bibfield  {author} {\bibinfo {author} {\bibfnamefont {F.}~\bibnamefont
  {Paillusson}} and\
  \bibinfo {author} {\bibfnamefont {R.}~\bibnamefont {Blossey}},\ }\href
  {\doibase https://doi.org/10.1103/PhysRevE.82.052501} {\bibfield
  {journal} {\bibinfo  {journal} {Phys. Rev. E}\ }\textbf {\bibinfo
  {volume} {82}},\ \bibinfo {pages} {052501} (\bibinfo {year}
  {2010})}\BibitemShut {NoStop}%
\bibitem [{\citenamefont {Berthoumieux}\ \emph {et~al.}(2019)\citenamefont {Berthoumieux},
  \citenamefont {Berthoumieux} and\ \citenamefont
  {Paillusson}}]{Berthoumieux2019}%
  \BibitemOpen
  \bibfield  {author} {\bibinfo {author} {\bibfnamefont {H.}~\bibnamefont
  {Berthoumieux}} and\
  \bibinfo {author} {\bibfnamefont {F.}~\bibnamefont
  {Paillusson}},\ }\href
  {\doibase https://doi.org/10.1063/1.5080183} {\bibfield
  {journal} {\bibinfo  {journal} {J. Chem. Phys.}\ }\textbf {\bibinfo
  {volume} {150}},\ \bibinfo {pages} {094507} (\bibinfo {year}
  {2019})}\BibitemShut {NoStop}%
\bibitem [{\citenamefont {Blossey}\ \emph {et~al.}(2022)\citenamefont {Blossey},
  \citenamefont {Blossey} and\ \citenamefont
  {Podgornik}}]{Blossey2022}%
  \BibitemOpen
  \bibfield  {author} {\bibinfo {author} {\bibfnamefont {R.}~\bibnamefont
  {Blossey}} and\
  \bibinfo {author} {\bibfnamefont {R.}~\bibnamefont
  {Podgornik}},\ }\href
  {\doibase https://doi.org/10.1209/0295-5075/ac7d0a} {\bibfield
  {journal} {\bibinfo  {journal} {Europhys. Lett.}\ }\textbf {\bibinfo
  {volume} {139}},\ \bibinfo {pages} {27002} (\bibinfo {year}
  {2022})}\BibitemShut {NoStop}%
\bibitem [{\citenamefont {Thompson}\ \emph {et~al.}(2022)\citenamefont
  {Thompson}, \citenamefont {Aktulga}, \citenamefont {Berger}, \citenamefont
  {Bolintineanu}, \citenamefont {Brown}, \citenamefont {Crozier}, \citenamefont
  {in~'t Veld}, \citenamefont {Kohlmeyer}, \citenamefont {Moore}, \citenamefont
  {Nguyen}, \citenamefont {Shan}, \citenamefont {Stevens}, \citenamefont
  {Tranchida}, \citenamefont {Trott},\ and\ \citenamefont
  {Plimpton}}]{Thompson2022}%
  \BibitemOpen
  \bibfield  {author} {\bibinfo {author} {\bibfnamefont {A.~P.}\ \bibnamefont
  {Thompson}}, \bibinfo {author} {\bibfnamefont {H.~M.}\ \bibnamefont
  {Aktulga}}, \bibinfo {author} {\bibfnamefont {R.}~\bibnamefont {Berger}},
  \bibinfo {author} {\bibfnamefont {D.~S.}\ \bibnamefont {Bolintineanu}},
  \bibinfo {author} {\bibfnamefont {W.~M.}\ \bibnamefont {Brown}}, \bibinfo
  {author} {\bibfnamefont {P.~S.}\ \bibnamefont {Crozier}}, \bibinfo {author}
  {\bibfnamefont {P.~J.}\ \bibnamefont {in~'t Veld}}, \bibinfo {author}
  {\bibfnamefont {A.}~\bibnamefont {Kohlmeyer}}, \bibinfo {author}
  {\bibfnamefont {S.~G.}\ \bibnamefont {Moore}}, \bibinfo {author}
  {\bibfnamefont {T.~D.}\ \bibnamefont {Nguyen}}, \bibinfo {author}
  {\bibfnamefont {R.}~\bibnamefont {Shan}}, \bibinfo {author} {\bibfnamefont
  {M.~J.}\ \bibnamefont {Stevens}}, \bibinfo {author} {\bibfnamefont
  {J.}~\bibnamefont {Tranchida}}, \bibinfo {author} {\bibfnamefont
  {C.}~\bibnamefont {Trott}}, \ and\ \bibinfo {author} {\bibfnamefont {S.~J.}\
  \bibnamefont {Plimpton}},\ }\href {\doibase 10.1016/J.CPC.2021.108171}
  {\bibfield  {journal} {\bibinfo  {journal} {Comput. Phys. Commun.}\ }\textbf
  {\bibinfo {volume} {271}},\ \bibinfo {pages} {108171} (\bibinfo {year}
  {2022})}\BibitemShut {NoStop}%
\bibitem [{\citenamefont {Xu}\ \emph {et~al.}(2013)\citenamefont {Xu},
  \citenamefont {Liang}, \citenamefont {Shi},\ and\ \citenamefont
  {Chen}}]{Xu2013}%
  \BibitemOpen
  \bibfield  {author} {\bibinfo {author} {\bibfnamefont {M.}~\bibnamefont
  {Xu}}, \bibinfo {author} {\bibfnamefont {T.}~\bibnamefont {Liang}}, \bibinfo
  {author} {\bibfnamefont {M.}~\bibnamefont {Shi}}, \ and\ \bibinfo {author}
  {\bibfnamefont {H.}~\bibnamefont {Chen}},\ }\href {\doibase
  10.1021/CR300263A} {\bibfield  {journal} {\bibinfo  {journal} {Chem. Rev.}\
  }\textbf {\bibinfo {volume} {113}},\ \bibinfo {pages} {3766} (\bibinfo {year}
  {2013})}\BibitemShut {NoStop}%
  \bibitem [{\citenamefont {Kumar}\ \emph {et~al.}(2007)\citenamefont {Kumar},
  \citenamefont {Starr}, \citenamefont {Buldyrev},\ and\ \citenamefont
  {Stanley}}]{Kumar2007}%
  \BibitemOpen
  \bibfield  {author} {\bibinfo {author} {\bibfnamefont {P.}~\bibnamefont
  {Kumar}}, \bibinfo {author} {\bibfnamefont {F.W.}~\bibnamefont {Starr}}, \bibinfo
  {author} {\bibfnamefont {S.V.}~\bibnamefont {Buldyrev}}, \ and\ \bibinfo {author}
  {\bibfnamefont {H.E.}~\bibnamefont {Stanley}},\ }\href {\doibase
  10.1103/PhysRevE.75.011202} {\bibfield  {journal} {\bibinfo  {journal} {Phys. Rev. E}\
  }\textbf {\bibinfo {volume} {75}},\ \bibinfo {pages} {011202} (\bibinfo {year}
  {2007})}\BibitemShut {NoStop}%
\bibitem [{\citenamefont {Good}\ and\ \citenamefont {Hope}(1970)}]{Good1970}%
  \BibitemOpen
  \bibfield  {author} {\bibinfo {author} {\bibfnamefont {R.~J.}\ \bibnamefont
  {Good}}\ and\ \bibinfo {author} {\bibfnamefont {C.~J.}\ \bibnamefont
  {Hope}},\ }\href {\doibase https://doi.org/10.1063/1.1674022} {\bibfield
  {journal} {\bibinfo  {journal} {J. Chem. Phys.}\ }\textbf {\bibinfo {volume}
  {53}},\ \bibinfo {pages} {540} (\bibinfo {year} {1970})}\BibitemShut
  {NoStop}%
\bibitem [{\citenamefont {Whitley}\ and\ \citenamefont {Smith}(2004)}]{Whitley2004}%
  \BibitemOpen
  \bibfield  {author} {\bibinfo {author} {\bibfnamefont {H.~D.}\ \bibnamefont
  {Whitley}}\ and\ \bibinfo {author} {\bibfnamefont {D.~E.}\ \bibnamefont
  {Smith}},\ }\href {\doibase https://doi.org/10.1063/1.1648013} {\bibfield
  {journal} {\bibinfo  {journal} {J. Chem. Phys.}\ }\textbf {\bibinfo {volume}
  {120}},\ \bibinfo {pages} {5387} (\bibinfo {year} {2004})}\BibitemShut
  {NoStop}%
\bibitem [{\citenamefont {Zhu}\ \emph {et~al.}(2002)\citenamefont {Zhu},
  \citenamefont {Tajkhorshid},\ and\ \citenamefont {Schulten}}]{Zhu2002}%
  \BibitemOpen
  \bibfield  {author} {\bibinfo {author} {\bibfnamefont {F.}~\bibnamefont
  {Zhu}}, \bibinfo {author} {\bibfnamefont {E.}~\bibnamefont {Tajkhorshid}}, \
  and\ \bibinfo {author} {\bibfnamefont {K.}~\bibnamefont {Schulten}},\ }\href
  {\doibase https://doi.org/10.1016/S0006-3495(02)75157-6} {\bibfield
  {journal} {\bibinfo  {journal} {Biophys. J.}\ }\textbf {\bibinfo {volume}
  {83}},\ \bibinfo {pages} {154} (\bibinfo {year} {2002})}\BibitemShut
  {NoStop}%
  \bibitem [{\citenamefont {COMSOL}\ \emph {et~al.}(2018)\citenamefont {COMSOL}}]{COMSOL}%
  \BibitemOpen
  \bibfield  {author} {\bibinfo {author} {\bibnamefont
  {COMSOL Multiphysics \textregistered~v. 5.4, www.comsol.com. COMSOL AB, Stockholm, Sweden}} \
  }\ \BibitemShut
  {NoStop}%
\bibitem [{\citenamefont {Fragasso}\ \emph {et~al.}(2019)\citenamefont
  {Fragasso}, \citenamefont {Pud},\ and\ \citenamefont
  {Dekker}}]{Fragasso2019}%
  \BibitemOpen
  \bibfield  {author} {\bibinfo {author} {\bibfnamefont {A.}~\bibnamefont
  {Fragasso}}, \bibinfo {author} {\bibfnamefont {S.}~\bibnamefont {Pud}}, \
  and\ \bibinfo {author} {\bibfnamefont {C.}~\bibnamefont {Dekker}},\ }\href
  {\doibase https://doi.org/10.1088/1361-6528/ab2d35} {\bibfield  {journal}
  {\bibinfo  {journal} {Nanotechnology}\ }\textbf {\bibinfo {volume} {30}},\
  \bibinfo {pages} {395202} (\bibinfo {year} {2019})}\BibitemShut {NoStop}%
\bibitem [{\citenamefont {Heerema}\ \emph {et~al.}(2015)\citenamefont
  {Heerema}, \citenamefont {Schneider}, \citenamefont {Rozemuller},
  \citenamefont {Vicarelli}, \citenamefont {Zandbergen},\ and\ \citenamefont
  {Dekker}}]{Heerema2015}%
  \BibitemOpen
  \bibfield  {author} {\bibinfo {author} {\bibfnamefont {S.}~\bibnamefont
  {Heerema}}, \bibinfo {author} {\bibfnamefont {G.}~\bibnamefont {Schneider}},
  \bibinfo {author} {\bibfnamefont {M.}~\bibnamefont {Rozemuller}}, \bibinfo
  {author} {\bibfnamefont {L.}~\bibnamefont {Vicarelli}}, \bibinfo {author}
  {\bibfnamefont {H.}~\bibnamefont {Zandbergen}}, \ and\ \bibinfo {author}
  {\bibfnamefont {C.}~\bibnamefont {Dekker}},\ }\href {\doibase
  https://doi.org/10.1088/1361-6528/ab2d35} {\bibfield  {journal} {\bibinfo
  {journal} {Nanotechnology}\ }\textbf {\bibinfo {volume} {26}},\ \bibinfo
  {pages} {074001} (\bibinfo {year} {2015})}\BibitemShut {NoStop}%
\bibitem [{\citenamefont {Chuang}\ \emph {et~al.}(2024)\citenamefont {Chuang},
  \citenamefont {Sowa}, \citenamefont {Kitazumi},\ and\ \citenamefont
  {Shirai}}]{Chuang2024}%
  \BibitemOpen
  \bibfield  {author} {\bibinfo {author} {\bibfnamefont {W.}~\bibnamefont
  {Chuang}}, \bibinfo {author} {\bibfnamefont {K.}~\bibnamefont {Sowa}},
  \bibinfo {author} {\bibfnamefont {Y.}~\bibnamefont {Kitazumi}}, \ and\
  \bibinfo {author} {\bibfnamefont {O.}~\bibnamefont {Shirai}},\ }\href
  {\doibase https://doi.org/10.1016/j.electacta.2024.144488} {\bibfield
  {journal} {\bibinfo  {journal} {Electrochim. Acta}\ }\textbf {\bibinfo
  {volume} {497}},\ \bibinfo {pages} {144488} (\bibinfo {year}
  {2024})}\BibitemShut {NoStop}%
  
\end{thebibliography}

\end{document}